\documentclass[12pt]{article}

\usepackage[table]{xcolor}
\usepackage[normalem]{ulem}
\usepackage{booktabs}
\usepackage{longtable}
\usepackage{footmisc}

\usepackage{scicite}
\usepackage{ref_macros_sci}

\usepackage{cancel}
\usepackage[normalem]{ulem}
\usepackage{graphicx}
\usepackage{setspace}
\usepackage[hyphens]{url}
\usepackage{graphicx}
\usepackage{amsmath}
\usepackage{amssymb}
\usepackage{xspace}
\usepackage{multirow}
\usepackage{array}
\newcolumntype{C}[1]{>{\centering\arraybackslash}p{#1}}
\usepackage{rotating}

\newcommand\T{\rule{0pt}{4.ex}}       

\usepackage[final]{pdfpages}

\usepackage{times}

\newenvironment{sciabstract}{%
\begin{quote} \bf}
{\end{quote}}

\newcounter{lastnote}
\newenvironment{scilastnote}{%
\setcounter{lastnote}{\value{enumiv}}%
\addtocounter{lastnote}{+1}%
\begin{list}%
{\arabic{lastnote}.}
{\setlength{\leftmargin}{.22in}}
{\setlength{\labelsep}{.5em}}}
{\end{list}}

\title{Evidence for a Compact Object in the Aftermath of the Extra-Galactic Transient AT2018cow}
\author
{Dheeraj R. Pasham$^{1\ast}$,
Wynn C.~G. Ho$^{2}$,
William Alston$^{3}$,
Ronald Remillard$^{1}$,\\
Mason Ng$^{1}$,
Keith Gendreau$^{4}$,
Brian D.~Metzger$^{5}$, 
Diego Altamirano$^{6}$,\\
Deepto Chakrabarty$^{1}$,
Andrew Fabian$^{7}$,
Jon Miller$^{8}$,
Peter Bult$^{4,9}$, \\
Zaven Arzoumanian$^{4}$,
James F. Steiner$^{10}$ 
Tod Strohmayer$^{4}$,\\ 
Francesco Tombesi$^{4,9,11,12}$,
Jeroen Homan$^{13}$, 
Edward M. Cackett$^{14}$,
Alice Harding$^{15}$\\
\\
\normalsize{$^{1}$Kavli Institute for Astrophysics and Space Research, Massachusetts Institute of Technology,}\\
\normalsize{Cambridge, MA 02139, USA}\\
\normalsize{$^{2}$Department of Physics and Astronomy, Haverford College, 370 Lancaster Avenue,} \\\normalsize{Haverford, PA 19041, USA}\\
\normalsize{$^{3}$European Space Agency (ESA), European Space Astronomy Centre (ESAC),} \\\normalsize{ Villanueva de la Ca\~{n}ada, Madrid, E-28691, Spain}\\ 
\normalsize{$^{4}$NASA Goddard Space Flight Center, Greenbelt, MD, USA}\\
\normalsize{$^{5}$Columbia Astrophysics Laboratory, Columbia University, New York, NY 10027, USA}\\
\normalsize{$^{6}$Department of Physics and astronomy, University of Southampton, }\\\normalsize{University Road, Southampton, SO17 1BJ, UK}\\
\normalsize{$^{7}$Institute of Astronomy, University of Cambridge, UK}\\
\normalsize{$^{8}$Department of Astronomy, 311 West Hall, 1085 South University Ave.,}\\ \normalsize{Ann Arbor, MI 48109-1107}\\
\normalsize{$^{9}$Department of astronomy, University of Maryland College Park, MD, 20742}\\
\normalsize{$^{10}$Center for Astrophysics, Harvard \& Smithsonian, Cambridge, MA, 02138}\\
\normalsize{$^{11}$Department of Physics, Tor Vergata University of Rome, Via della Ricerca Scientifica 1},\\ \normalsize{00133 Rome, Italy}\\
\normalsize{$^{12}$INAF – Astronomical Observatory of Rome, Via Frascati 33, 00040 Monte Porzio Catone, Italy}\\
\normalsize{$^{13}$Eureka Scientific, Inc., Oakland, CA 94602, USA}\\
\normalsize{$^{14}$Department of Physics and Astronomy, Wayne State University, 666 W Hancock,}\\ \normalsize{Detroit, MI 48201}\\
\normalsize{$^{15}$Theoretical Division, Los Alamos National Laboratory, Los Alamos, NM 87545}\\
\\
\normalsize{$^\ast$To whom correspondence should be addressed; E-mail:  dheeraj@space.mit.edu}
}

\date{}

\begin{document}

\baselineskip24pt
\maketitle 

\begin{sciabstract}
The brightest Fast Blue Optical Transients (FBOTs) are mysterious extragalactic explosions that may represent a new class of astrophysical phenomena \cite{FBOTReview}.  Their fast time to maximum brightness of less than a week and decline over several months and atypical optical spectra and evolution are difficult to explain within the context of core-collapse of massive stars which are powered by radioactive decay of Nickel-56 and evolve more slowly \cite{Drout14,DarkenergyFBOTs}.  AT2018cow (at redshift of 0.014) is an extreme FBOT in terms of rapid evolution and high luminosities \cite{rivera19,Raf19,Prentice19,Perley19}.  Here we present evidence for a high-amplitude quasi-periodic oscillation (QPO) of AT2018cow's soft X-rays with a frequency of 224 Hz (at 3.7$\sigma$ significance level or false alarm probability of 0.02\%) and fractional root-mean-squared amplitude of $>$30\%.  This signal is found in the average power density spectrum taken over the entire 60-day outburst and suggests a highly persistent signal that lasts for a billion cycles.  The high frequency (rapid timescale) of 224 Hz (4.4 ms) argues for a compact object in AT2018cow, which can be a neutron star or black hole with a mass less than 850 solar masses.  If the QPO is the spin period of a neutron star, we can set limits on the star's magnetic field strength.  Our work highlights a new way of using high time-resolution X-ray observations to study FBOTs.
\end{sciabstract}

High-cadence sky surveys that can scan the same portions of the sky multiple times per night have uncovered fast-evolving optical transients \cite{Drout14, DarkenergyFBOTs, SubaruFBOTs}. These ''fast'' transients rise to their peak brightness within $\lesssim$ 10 days and fade away within a month or two (e.g., \cite{Drout14}). They are spatially coincident with external galaxies but are offset from their nuclei (e.g., \cite{Rest18, Drout14}). Their optical spectra are often blue with occasional presence of Hydrogen and Helium features. The peak luminosities of FBOTs range from the faint end of core-collapse supernovae to the bright end of superluminous supernovae (see Fig. 1 of \cite{FBOTReview} and Fig. 1 of \cite{ztffbots}).  A recent study \cite{ztffbots} has shown that the majority of FBOTs are extreme cases of core-collapse supernovae. However, a subset of high luminosity FBOTs with peak bolometric luminosities $\sim$10$^{44}$ erg s$^{-1}$ cannot be explained as an extension of properties of core-collapse supernovae\cite{camel,cssfbot, koala}.  Thus, several alternate mechanisms mentioned above have been proposed to explain the properties of luminous FBOTs. These include emission from the interaction of the supernova shockwave with a dense circumstellar medium \cite{Raf19,cowcsm,fox}, injection of energy from spin down of a young magnetar formed either in a core-collapse supernova or a binary neutron star merger \cite{kasen10,metzger14}, accretion onto a newly formed compact object in a failed supernova \cite{Raf19}, mergers of binary white dwarfs \cite{cow_wdmerger}, and intermediate-mass black holes (IMBHs: a few$\times$10$^{4-5}$ $M_{\odot}$) tidally disrupting stars \cite{kuin19,Perley19}. 

Prior to June 2018,  the majority of FBOTs were first identified in archival images (e.g., \cite{Rest18, Drout14, tanakafbots, pursianen}). AT2018cow, discovered by the ATLAS sky survey\cite{atlas} in a galaxy at a distance of $\approx$ 60 Mpc \cite{Prentice19}, was discovered in real time. Its brightness rise of more than 5.7 magnitudes in just 4 days (see Fig. 1 of \cite{Prentice19}) was remarkable and its peak bolometric luminosity of 4$\times$10$^{44}$ erg s$^{-1}$ makes it the brightest FBOT  known so far \cite{FBOTReview}. As the discovery was promptly reported \cite{cowatel}, the source received a significant amount of multi-wavelength coverage. Radio, millimeter, optical, UV, X-ray and gamma ray properties of the source are described in various papers \cite{Raf19, Perley19,Prentice19, rivera19, kuin19, cowmm, almacow}. However, in spite of exquisite coverage the physical origin of AT2018cow remains elusive. 

Given its high X-ray luminosity (peak value of a roughly 10$^{43}$ erg/s, \cite{Raf19}) and variability on timescales of a few tens of hours \cite{Raf19, cowmm, rivera19}, compact object (accretion) powered scenarios have been proposed for AT2018cow. These suggestions include emission from tidal disruption of a star by an intermediate-mass black hole with mass in the range of 10$^{4-6}$ $M_{\odot}$ \cite{Perley19}, fallback accretion in a failed supernova \cite{Raf19, failedsne}, and energy injection by a newly born neutron star in a supernova \cite{cowmagnetar}. 

\begin{sloppypar}
Several works over the last few decades \cite{mcclin} have found that when accreting compact object (stellar-mass black hole/neutron star) X-ray binaries go into outbursts--due to enhanced accretion--they sometimes exhibit high-frequency (a few$\times$(10-100) Hz) quasi-periodic variability in their X-ray brightness. There is no clear consensus  on the exact mechanisms that produce these so-called High-Frequency Quasi-Periodic Oscillations (HFQPOs) but it is generally agreed upon that they originate from a region close to the compact object where the dynamics of motion are dictated by the compact object's strong gravitational field (see \cite{mcclin} and references therein), and they represent a direct evidence for the presence of a compact object. Some active galactic nuclei have also shown evidence for QPOs which have been argued to be analogous to HFQPOs of stellar-mass black holes\cite{alstonqpo,agnqpocatalog,rejqpo}. More recently, HFQPO analogs  (frequencies of a few mHz) have also been found in stellar tidal disruption events involving $\sim$10$^{6}$ $M_{\odot}$ black holes \cite{14liqpo, 1644qpo, dachengqpo} suggesting that, perhaps, such QPOs are universal among all compact object systems that undergo extreme changes in accretion.

\end{sloppypar}

\begin{figure}[ht]
\begin{center}
\includegraphics[width=6.25in, angle=0]{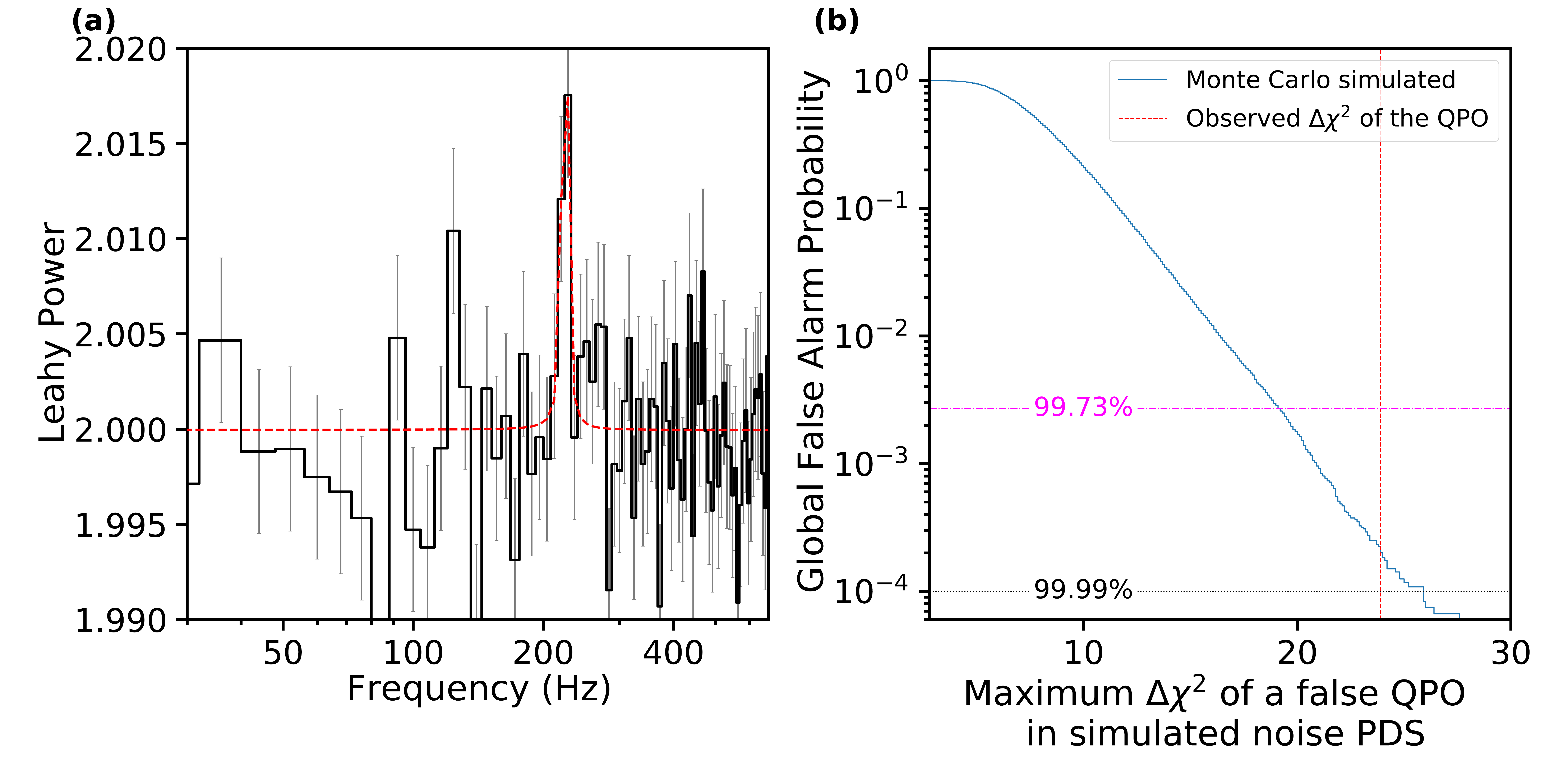}
\end{center}
\caption{\textbf{(a) Average X-ray PDS of AT2018cow showing evidence for a quasi-periodicity near 224 Hz.} This PDS was computed by averaging 105 256-second soft X-ray (0.25-2.5 keV) light curves sampled at 1/2048 s. The resulting PDS was further re-binned by a factor of 2048 which gives a frequency resolution of 8 Hz. The strongest excess above the Poisson noise level of 2 is around 224 Hz. The power values in the rest of the PDS continuum are consistent with white noise (see SM  sections \ref{supsec:kstest} and \ref{supsec:fap} and Figures \ref{fig:kstest}, \ref{fig:probplot}, \ref{fig:xmmpds}). The PDS is normalized such that the mean value surrounding 224 Hz is equal to the Poisson value of 2. The best-fit constant + Lorentzian models are indicated by the dashed red curve. \textbf{(b) Estimates for the statistical significance of the 224 Hz QPO.} The likelihood of finding a QPO from noise fluctuations{\bf, i.e., false alarm probability = 1-CDF,} (y-axis) vs the maximum improvement in $\chi^2$ by fitting the simulated noise PDS with a constant+Lorentzian over modeling it with a constant at every frequency searched during the identification of the signal on the left (see section \ref{supsec:mcsims} and Fig. \ref{fig:pdcdfs} for more details). {\bf The global false alarm probability of finding a QPO as strong as the one observed is $\approx$ 2$\times$10$^{-4}$ (1 in 5000; $\approx$3.7$\sigma$)}. \label{fig:fig1}}
\end{figure}

To test these hypotheses for a compact object (accretion) powered scenario, we studied AT2018cow's X-ray (0.25-2.5 keV) variability using an average power density spectrum derived from the entire monitoring data taken by the Neutron Star Interior Composition Explorer ({\it NICER}) on board the International Space Station. We find evidence for an X-ray QPO in the average PDS (left panel of Fig. \ref{fig:fig1}). The QPO signal has a centroid frequency, full width half maximum (FWHM) and a fractional root-mean-squared amplitude of 224.4$\pm$1.0 Hz, $<$ 16 Hz, and 30$\pm$3\%, respectively. Using a rigorous Monte Carlo approach (see supplementary material, SM) that takes into account the nature of the underlying noise continuum and the search trials, we find the global false alarm probability of this signal to be $\approx$2$\times$10$^{-4}$ (or 3.7$\sigma$ equivalent for a normal distribution; see right panel of Fig. \ref{fig:fig1} and SM). 

\begin{figure}[ht]
\begin{center}
\includegraphics[width=0.6\textwidth, angle=0]{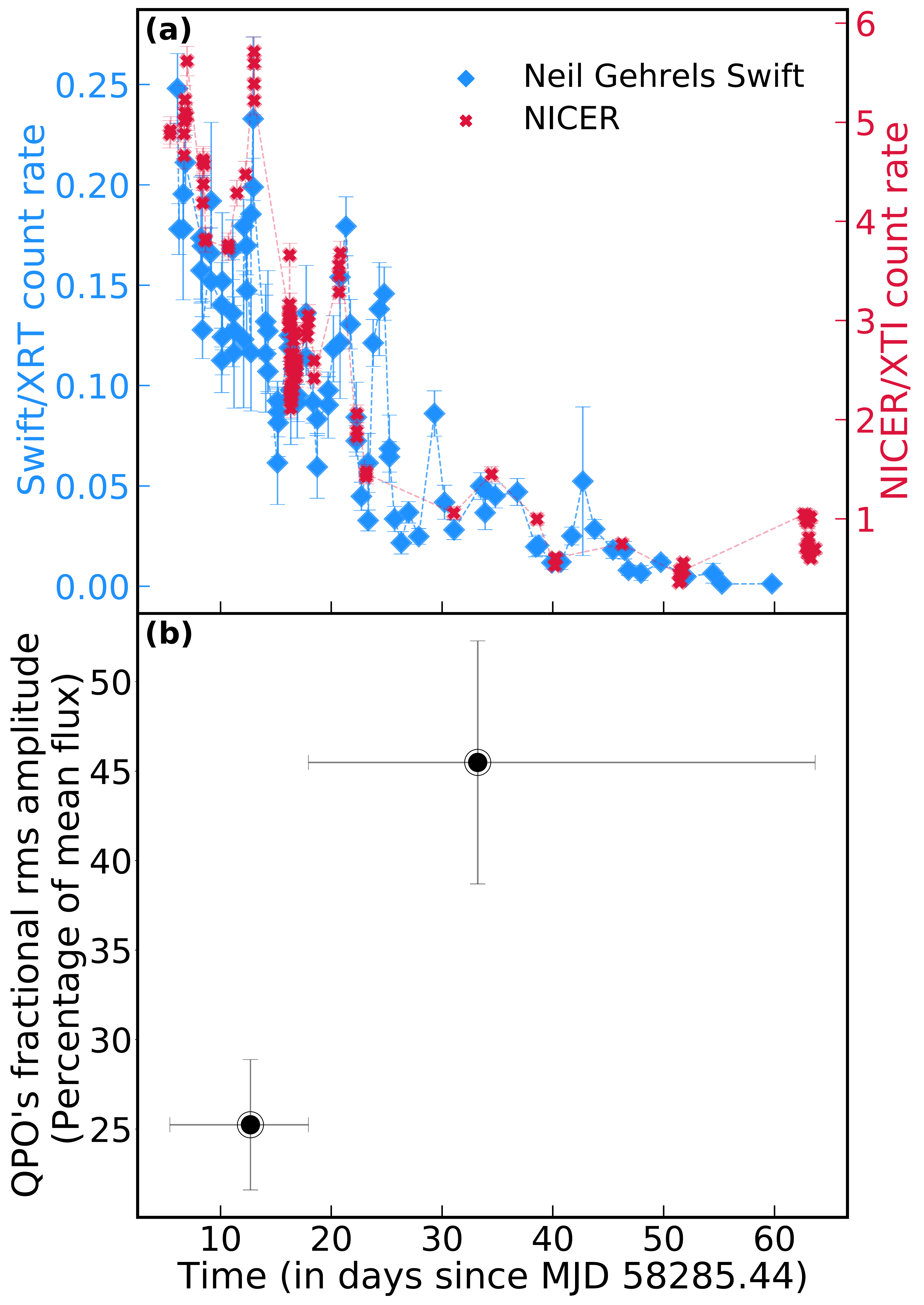}
\end{center}
\caption{{\bf (a) Comparison of {\it NICER}/XTI and Neil Gehrels {\it Swift}/XRT long-term light curves.} Both y-axes are in units of counts/sec. It is evident that AT2018cow's long-term soft X-ray (0.25-2.5 keV) variability as observed by the non-imaging {\it NICER} telescope is same as that observed with Neil Gehrels {\it Swift}/XRT (0.3-2.5 keV) which has imaging capability. This suggests that the flux observed by {\it NICER} is dominated by AT2018cow with minimal contamination from other nearby astrophysical sources. {\bf (b) Fractional root-mean-squared amplitude of the 224 Hz QPO vs time.} The signal appears to be stronger during the end of the outburst. The two values are 25$\pm$4 and 45$\pm$7\% corresponding to exposures during 12.7$^{+5.2}_{-7.3}$ and 33.2$^{+30.5}_{-15.3}$ days, respectively. The jump in the QPO's strength coincides with higher levels of variability on days timescale (XRT light curve in (a)) and also with the dramatic changes in optical spectra seen around the same time\cite{Raf19}.}\label{fig:fig2}
\end{figure}

\begin{figure}[ht]
\begin{center}
\includegraphics[width=\textwidth, angle=0]{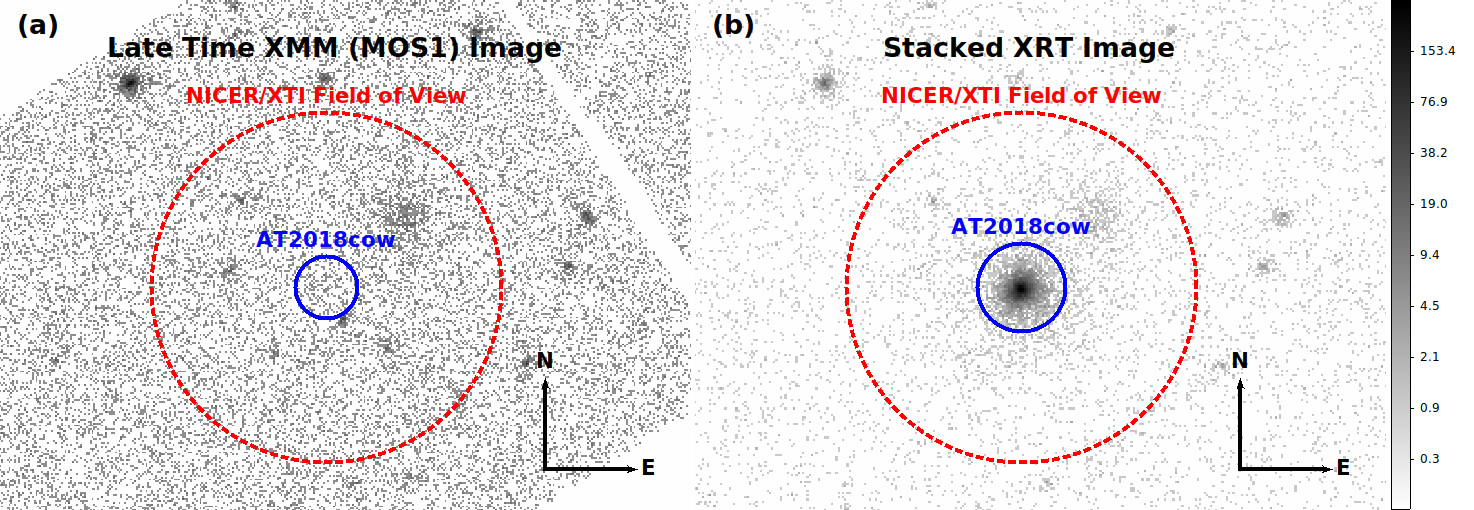}
\end{center}
\caption{{\bf {\it XMM-Newton} and {\it Swift} images of {\it NICER}'s field of view of AT2018cow showing that there was minimal contamination from field sources. (a) {\it XMM-Newton}/MOS1 image of AT2018cow's field of view long after AT2018cow faded.} The position of AT2018cow is indicated by a blue circle with a radius of 33''. There is no contaminating point source directly coincident with AT2018cow. {\bf (b) Stacked Neil Gehrels {\it Swift}/XRT image of the field of view of AT2018cow.} The blue circle of radius 47'' is centered on AT2018cow's optical position (16:16:00.220 +22:16:04.91; J2000.0). This particular image was extracted by using all the archival XRT images as of July 2020. It is evident that while there are a few point sources present in {\it NICER}'s field of view, their contribution to overall XTI flux is negligible when compared to AT2018cow (see also Fig. \ref{fig:fig2}). In both panels the outer/red (dashed) circles show {\it NICER}/XTI's approximate field of view of 3.1'. The north and east arrows are each 100'' long. The colorbar show counts. }\label{fig:fig3}
\end{figure}

We rule out various instrumental and particle backgrounds as the origin for this QPO signal (See SM). Based on the long term light curves derived from {\it NICER} and Neil Gehrels {\it Swift} X-ray telescope, which has imaging capability (see top panel of Fig. \ref{fig:fig2}), and late time {\it XMM-Newton} X-ray images of AT2018cow's field of view (FoV), we also rule out a contaminating source as the origin of this QPO (see Fig. \ref{fig:fig3}). Furthermore, over the last three years of {\it NICER} operations a signal of similar nature has never been found in any of the several dozens of other targets (see, for example, Fig. \ref{fig:agnpds}). Based on these tests, we conclude that the signal is consistent with originating from AT2018cow.

After establishing that the signal is statistically significant and ruling out an instrumental and a background origin, we extracted the QPO's signal-to-noise as a function of the accumulated exposure (See Fig. \ref{fig:deltachi} and supplementary movie S1). It is evident that the QPO's strength increases gradually with increasing exposure. This suggests that the signal is persistent and present at some level throughout the $\sim$ 2 month monitoring period. Taken at face value, this suggests that it is stable over $\sim$60 days/4.44 ms  $\gtrsim$ 10$^{9}$ cycles. Interestingly, the mean slope of the curve in Fig. \ref{fig:deltachi} also steepens around the same time (near day 17) when high-amplitude X-ray flares start to appear on days timescale (see the blue diamonds in the upper panel of Fig. \ref{fig:fig2}). Moreover, the fractional rms of the QPO jumps around day 17 which also coincides with the time when the optical spectrum of AT2018cow underwent dramatic changes (see \cite{Raf19} for details). We separated the total exposure into two epochs, before and after day 17.  We extract an average PDS from each of these two time intervals and the QPO is fit with a Lorentzian. The QPO's fractional rms amplitude appears to be higher at later times (see bottom panel of Fig. \ref{fig:fig2}).

The frequency of this QPO alone can set stringent constraints on the underlying physical mechanism producing X-rays in AT2018cow. The causality argument suggests that the physical size of an object producing this signal cannot be larger than the light crossing size, i.e., speed of light $\times$ (1/224 s) $\approx$ 1.3$\times$10$^{8}$ cm. This small size points us naturally towards a compact object. Emission from shock interactions\cite{Raf19, cowcsm} is disfavored because CSM--shock interactions are not known to ''pulse''. If the compact object is a black hole then assuming the emission originates from the innermost stable circular orbit allows us to set an upper limit on the black hole mass. These limits are 95 $M_{\odot}$ and 850 $M_{\odot}$ for a maximally prograde spinning (spin has the same direction as the material falling in) and a retrograde (vice versa) spinning black hole, respectively. A larger emission radius (in units of gravitational radii) would require the object to be even more compact. This rules out a heavy IMBH ($\gtrsim$ 850 $M_{\odot}$) in AT2018cow \cite{kuin19}.

\begin{sloppypar}
HFQPOs with frequencies in the hundreds of Hz range have been seen in a handful of stellar-mass black holes with known dynamical mass estimates \cite{mcclin}. They often appear in pairs with frequency ratio of 2:3 \cite{mcclin}. In these systems it appears that the black hole mass scales inversely with the HFQPO frequency (see, for example, Fig. 4.17 of \cite{mcclin}). While the frequency and width of AT2018cow's QPO are similar to HFQPOs of stellar-mass black holes, the observed fractional rms amplitude is high (25-45\% compared to a few percent in stellar-mass black holes \cite{mcclin}), and a harmonic is not apparent here (see sec. \ref{supsec:harmonics} and Table \ref{tab:rmstab} for upper limits). Furthermore, the QPO's energy dependence is also uncertain given our narrow X-ray band pass of 0.25-2.5 keV. Also, the absence of any red noise (even at lower frequencies, i.e., a few$\times$mHz; see Fig. \ref{fig:xmmpds}) is unlike accreting X-ray binaries. All the stellar-mass X-ray binaries with HFQPOs are relatively highly inclined \cite{mcclin}. Given AT2018cow's high (a few$\times$10$^{42}$ erg/s) average X-ray luminosity beaming, which can boost the apparent brightness, is likely present, and this points towards a low inclination for this system. Thus, given all these factors, it is unclear if a direct comparison can be made with HFQPOs of stellar-mass black hole binaries. Nevertheless, if we assume the same scaling law applies and that 224 Hz is the fundamental harmonic, then the implied black hole mass in AT2018cow is $\sim$ 4 $M_{\odot}$.
\end{sloppypar}

Alternatively, the compact object could be a neutron star with the QPO representing either its spin rate of 224~Hz or an analog of the upper kHz QPOs which can have high fractional rms amplitudes (see Fig. 2 of \cite{khzqpos}). We explore multiple physical scenarios for the former case. First, we consider the case where AT2018cow's luminosity is driven by the spin down of an isolated millisecond magnetar. In this case, the high and changing bolometric luminosity would require high magnetic field strength which in turn would lead to a rapid change in the spin frequency. This is inconsistent with the QPO's stability based on its width ($<16\mbox{ Hz}/60\mbox{ d}=3\times 10^{-6}\mbox{ Hz s$^{-1}$}$). This inconsistency rules out an isolated magnetar powered scenario. Then we consider a case in which the luminosity is powered by accretion of matter onto the magnetic poles of a neutron star. While this model provides a possible mechanism to induce periodicity in the light curve and a declining luminosity, its prediction for spin period evolution is inconsistent with the QPO's stability (see SM for more details). Lastly, if this QPO is due to a process similar to that operating in magnetar QPOs, the physics and damping mechanisms would be in an entirely different regime given that the stability of AT2018cow's QPO is significantly ($\gtrsim$10$^{5}$) longer than those of magnetar burst QPOs\cite{magbursts}.


Within the bounds of a central engine being either a neutron star, stellar-mass black hole or a few hundred M$_{\odot}$ black hole, the QPO presented here can be produced under different scenarios. For example, it has been proposed that a flare like AT2018cow could arise from young star clusters containing a stellar-mass black hole (mass $\sim$ 10 M$_{\odot}$) tidally disrupting a main sequence star  within the cluster\cite{kyle20} or from fallback accretion of material in a failed supernova whose core collapses into either a neutron star or a stellar-mass black hole \cite{Perley19}. In all these scenarios, such a QPO signal could be envisioned. Using the Atacama Large Millimeter/submillimeter Array (ALMA) \cite{almacow} spatially resolved the molecular gas content and star formation rate within AT2018cow's host galaxy. They find that AT2018cow is situated between a peak in molecular gas content and a blue star cluster. The presence of these regions indicate active star formation around AT2018cow and favors an association with deaths of massive stars for AT2018cow. The implications of this QPO for AT2018cow's existing models are summarized in Table \ref{table}.

In summary, we present evidence for a 224 Hz X-ray QPO persistent for one billion cycles from AT2018cow statistically significant at the 3.7$\sigma$ level. Previous works \cite{Raf19} have suggested a compact object in AT2018cow based on X-ray variability on a few tens of hours timescale and also its X-ray spectral resemblance to some accreting systems.  The detection of regular variations on millisecond timescale presented here provides the most direct way to infer the presence of a compact object, although it still remains unclear whether that compact object is a stellar-mass black hole or a neutron star. If AT2018cow originated from the death of a massive star, our findings represent the birth of a compact object in a supernova. Similar signals in future FBOTs could allow astronomers to study infant compact objects immediately after birth. Assuming all luminous FBOTs are AT2018cow-like which have an estimated volumetric rate of $<$10$^{-7}$ yr$^{-1}$ Mpc$^{-3}$\cite{ztffbots, ccrate1, ccrate2, cssfbot}, {\it NICER} can, in principle, find and study QPOs from one such system every 3 years (see SM sec. \ref{supsec:rates}).


\clearpage
\begin{table}
\footnotesize
    \centering
    \caption{Models for AT2018cow's X-ray emission and their validity against the 224 Hz QPO reported here. This table can be considered as a modified extension of Table 2 of \cite{Raf19}. }
    \label{table}
    \begin{tabular}{|C{0.25\textwidth}|C{0.11\textwidth}|C{0.11\textwidth}|C{0.455\textwidth}|}
        \hline
        \textbf{Model/class of models} & \textbf{References$^{\bullet}$} & \textbf{Consistent with QPO?} & \textbf{Notes} \T \\ \hline \T 
         Shock interactions with circumstellar medium (CSM) & \cite{cowcsm, rivera19} & No & CSM interaction could explain emission at other (non X-ray) wavelengths, i.e., optical and radio but inconsistent with rapid X-ray variability  \\ \T
         An embedded internal shock formed from interaction with dense CSM & \cite{Raf19} & No & A compact embedded internal shock is, in principle, consistent with the size constraint provided by the 224 Hz QPO. However, all X-ray QPOs known thus far in literature are from compact objects. So an internal shock model is disfavored.  \\ \T
         Accreting intermediate-mass black hole ($\gtrsim$10$^{3}$ $M_{\odot}$) & \cite{Perley19, imbhtde, kuin19} & No & Based on causality argument the compact object producing the QPO has to be less than 850 $M_{\odot}$ (see main text) \\ \T
         Neutron star (formed from merging white dwarfs/Supernova) & \cite{cow_wdmerger, Prentice19} & Yes & Constraints on magnetic field if the 224 Hz QPO represents the spin period. However, the persistence of the signal in a narrow frequency range is challenging to explain (see SM) \\ \T
         Stellar-mass black hole (accreting from outer layers of a failed supernova/tidally disrupting a star) & \cite{bhfallback, Perley19, kyle20, quatcow} & Yes &  The QPO frequency is similar to those often seen in known stellar-mass black holes but the  X-ray luminosity, rms, and QPO's stability are unlike any known stellar-mass black hole systems (see main text) \\
         &&& \\ \hline
    \end{tabular}
    $^{\bullet}$Not an exhaustive list. Also see references therein. 
\end{table}


\clearpage
\vfill\eject


\clearpage
\vfill\eject


\clearpage
\vfill\eject


\bibliographystyle{Science}
\bibliography{QPO_references}

\clearpage





\singlespace

\begin{scilastnote}
\item[] {\bf Supplementary Materials.}\\
Materials and Methods\\
Figs. S1 to S11\\
Table S1\\
Caption for movie S1\\
Movie S1\\
Supplementary Text\\
All the codes to download the data from HEASARC public archive, reduce it, and reproduce the results in this paper are available as supplementary files\\
\end{scilastnote}

%
%
\newpage
\newpage

\setcounter{page}{1}
\renewcommand{\theequation}{S\arabic{equation}}
\renewcommand{\thefigure}{S\arabic{figure}}
\renewcommand{\thetable}{S\arabic{table}}
\setcounter{figure}{0}

\singlespace


\section*{{\Huge Supplementary Material.}}


\section{\Large{\bf Data and Reduction. }}\label{supsec:data}
The primary data used in this study was acquired by {\it NICER}'s X-ray Timing Instrument (XTI; \cite{nicer}). We also utilized X-ray data from {\it Neil Gehrels Swift} observatory's \cite{swift} X-Ray Telescope (XRT; \cite{xrt}) and {\it XMM-Newton}'s \cite{xmm} European Photon Imaging Camera (EPIC; \cite{epicpn}). Below, we describe these data and their respective reduction methodologies in detail.


\subsection{{\it NICER}/XTI:}\label{supsec:nicerdatareduc} 
The {\it NICER} X-ray observatory on board the International Space Station (ISS) has been carrying out full science operations since July 2017. XTI, which operates in the 0.25-12 keV band, is the primary instrument on {\it NICER} and it consists of 56 co-aligned X-Ray Concentrators (XRC). Each XRC focuses X-rays into an aperture of a Focal Plane Module (FPM) which consists of a single pixel (non-imaging) Silicon Drift Detector (SDD; \cite{sdd}). At the beginning of science operations, 52 of the 56 FPMs were operational and together they provide an effective area of roughly 1900 cm$^{2}$ at 1.5 keV. This large effective area in the soft X-ray band combined with it's ability to provide an absolute time resolution of better than 300 nanoseconds makes {\it NICER} a unique facility to detect the fastest known soft X-ray astrophysical signals. 

\begin{sloppypar}

{\it NICER} started monitoring AT2018cow roughly five days after its discovery in the optical band on MJD 58285.44 \cite{cowatel, Prentice19}\footnote{Throughout this paper we refer times with respect to this optical discovery date}. In total, 26 sets of observations were made between MJD 58290.87 and 58349.11 with observation IDs running between 1200250101 and 1200250126. These datasets were publicly available and we downloaded them from the HEASARC  archive\footnote{\url{https://heasarc.gsfc.nasa.gov/cgi-bin/W3Browse/w3browse.pl}}. The cleaned events lists were extracted using the standard {\it NICER} Data Analysis Software (NICERDAS/HEASoft 6.28) tasks {\tt nicercal}, {\tt nimpumerge}, and {\tt nicerclean}. The latest NICER calibration release {\it xti20200722} (22 July 2020) was used. The cleaned event files were barycenter-corrected using the {\tt barycorr} {\tt ftools} task. AT2018cow's optical coordinates (J2000.0): (244.000917, +22.268031) were used along with {\it refframe=ICRS} and {\it ephem=JPLEPH.430}. The Good Time Intervals (GTIs) were extracted with the {\tt nimaketime} tool using the default filters: {\it nicersaafilt=YES}, {\it saafilt=NO}, {\it trackfilt=YES}, {\it ang\_dist=0.015}, {\it st\_valid=YES}, {\it elv=30}, {\it br\_earth=40}, {\it cor\_range="-"}, {\it min\_fpm=38}, {\it underonly\_range=0-200}, {\it overonly\_range=``0.0-1.0''}, {\it overonly\_expr=``1.52*COR\_SAX**(-0.633)''} Conservative {\it elv=30} and {\it br\_earth=40} were used to avoid optical loading by reflected light.  
\end{sloppypar}

\begin{figure}[ht]
\begin{center}
\includegraphics[width=6.0in, angle=0]{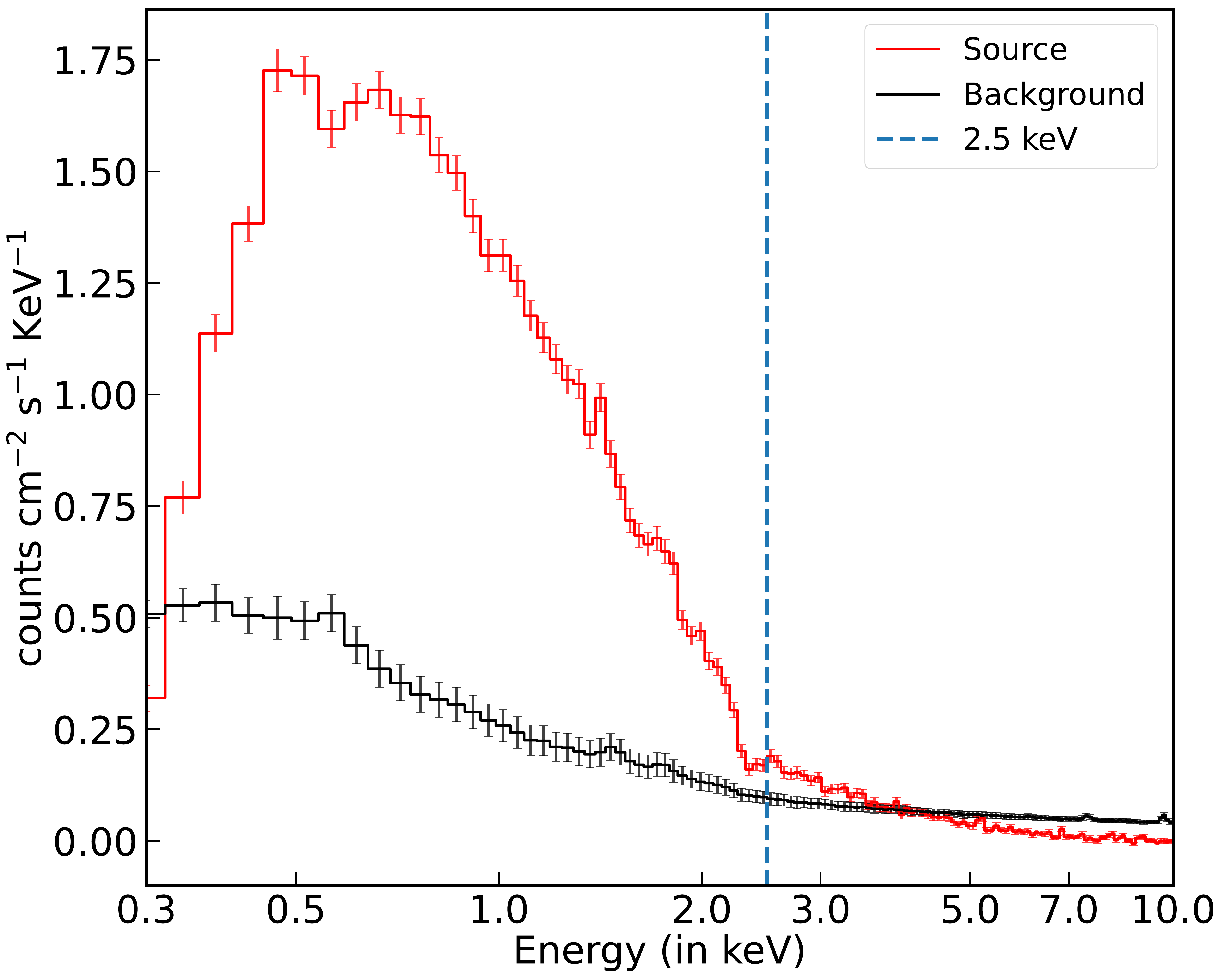}
\end{center}
\caption{{\textbf{Average energy spectra of AT2018cow (red) and the background (black).}} Beyond 2.5 keV the ratio of source to background counts (signal-to-noise) drops below 2. Because the sensitivity towards detecting a QPO depends on source$^{2}$/(source+background)\cite{vanderklis}, we chose to omit events above 2.5 keV.} \label{fig:espec}
\end{figure}

Due to a combination of decreasing effective area and increased contribution from high-energy particles the signal-to-noise (source-to-background) at energies $\gtrsim$ 2.5 keV deteriorates for faint X-ray targets like AT2018cow. To quantify this further we extracted an energy spectrum of AT2018cow using the entire {\it NICER} data along with an estimate of the average background spectrum using the 3c50 model\cite{3c50}. The spectrum in Fig. \ref{fig:espec} is binned using the optimal binning criterion of \cite{optmin} in addition to ensuring a minimum of 1 count per bin. It is evident that beyond 2.5 keV the ratio of source to background counts falls below 2. Therefore, to minimize the contribution from the highly variable particle background, which is particularly important for variability studies, we only considered X-ray events in the energy range of 0.25 to 2.5 keV. This background cannot be directly subtracted from the light curves, and thus acts to increase the noise in the power density spectra of the source. These are tested in detail in section \ref{supsec:instru}. Our entire download and reduction procedure can be easily replicated by running the codes that are available in supplementary files.


\subsection{Neil Gehrels {\it Swift}/XRT:}\label{supsec:xrtdatared}
Neil Gehrels {\it Swift} started monitoring AT2018cow on MJD 58288.44, roughly 3 days after its optical discovery. The observing cadence varied over the 58 d window coincident with the {\it NICER} observing campaign, i.e., until MJD 58349. For the first $\approx$3 weeks, Neil Gehrels {\it Swift} observed AT2018cow 3-4 times per day and thereafter the cadence was reduced to roughly one exposure per day. The individual exposure duration varied between 200 and 2000 s. In this work we only used XRT data in the band pass of 0.3-2.5 keV to be consistent with {\it NICER}'s energy range. 

We started our analysis with the publicly available, raw, level-1 data from HEASARC archive and reprocessed them with the {\tt xrtpipeline} task of HEASoft. Initially, when the source was bright, data was taken in both the Windowed Timing (WT) and the Photon Counting (PC) modes. But as the source count rate dropped observations were only carried out in the PC mode. For this work, we only used the PC mode data with event grades between 0 and 12. Source events were extracted from an annular region centered on AT2018cow with the outer radius fixed at 47''. This outer radius of 47'' corresponds to roughly 90\% (at 1.5 keV) of the light from a point source (as estimated from XRT's fractional encircled energy function). The inner radius was determined independently for each exposure by accounting for pile-up using the formalism described in the XRT user guide$\footnote{https://www.swift.ac.uk/analysis/xrt/pileup.php}$. Background events were extracted from an annular region centered on AT2018cow with an inner and outer radii of 150'' and 210'', respectively. These values were chosen to avoid any point sources in that background annulus.


\subsection{{\it XMM-Newton}:}\label{subsec:xmmdatared}
\begin{sloppypar}
{\it XMM-Newton} observed AT2018cow on three separate occasions roughly 37 (obsID: 0822580401; 33 ks), 82 (obsID: 0822580501; 45 ks), and 218 days (obsID: 0822580601; 56 ks) after its optical discovery. In this work, we only used the European Photon Imaging Camera (EPIC) data from the first and the last {\it XMM-Newton} epochs. The first exposure coincided with the NICER monitoring campaign while the last one was taken several months after AT2018cow faded away. As the data were taken in the {\it full frame mode} the Nyquist frequencies of the pn and MOS detectors were 6.8 Hz and roughly 0.2 Hz, respectively. To constrain the nature of variability on frequencies of a few Hz we only used pn data from the first dataset. For the last exposure MOS1 provides the best spatial resolution and hence we used only MOS1 data. 
\end{sloppypar}

We started {\it XMM-Newton}/EPIC data reduction with the raw Observation Data Files (ODF) and reprocessed them using the XMM Science Analysis Software's (xmmsas version 17.0.0) tasks {\tt epproc} and {\tt emproc} for the pn and MOS data, respectively. We employed standard data filters of {\it (PATTERN$\lesssim$12)} and {\it (PATTERN$\lesssim$4)} for the pn and the MOS data, respectively. We only considered events in the soft X-ray band of 0.25-2.5 keV to
be consistent with {\it NICER}'s data (see section \ref{supsec:nicerdatareduc}). We removed intervals of background flaring by manually inspecting the 10-12 keV light curve as outlined in the XMM-Newton data analysis guide. The source count rates were extracted from a circular region of radius 33'' which corresponds to 90\% of the light from a point source as estimated by the fractional encircled energy of EPIC instruments. Background events were extracted from two nearby circular regions of radius 50''.


\section{Analysis}\label{supsec: analysis}

\subsection{{\it NICER} identifies a QPO candidate}\label{supsec:qpodet}
We first divided the {\it NICER} data into 256-s continuous segments and extracted their light curves with a time resolution of 1/2048 s. This resulted in a total of 105 light curve segments, i.e., a cumulative exposure time of 26,880 s. With the mean count rate varying between 5.7 and 0.4 count s$^{-1}$ this choice of 256 s ensured $\gtrsim$ 100 counts in each segment. A Leahy normalized (Poisson noise level of 2) power density spectrum (PDS) was extracted from each of these individual light curves and they were all combined to obtain an average PDS (see the left panel of Fig. \ref{fig:fig1}). The PDS is consistent with a value of 2 (Poisson noise level) at all frequencies except for excess around 224 Hz. 


\subsection{Estimating the global statistical significance}\label{supsec:fap}
To estimate the global false alarm probability of this QPO candidate near 224 Hz by properly accounting for the underlying noise and also all the search trials we devised a Monte Carlo approach. The basic idea is to simulate a large number ($\sim$10$^{5}$ in our case) of random realizations of the underlying noise process, i.e., light curves that capture the underlying PDS continuum. Then, extract a power spectrum just like the real data, perform a global search (at all frequencies) for QPOs, and finally, estimate the probability of seeing a noise fluctuation that resembles a QPO as strong as the one found in the real data (left panel of Fig. \ref{fig:fig1}). Because a robust significance estimate relies on accurate knowledge of the underlying noise process, at first, we lay extra emphasis on understanding the nature of the PDS continuum. 

\subsubsection{Probability plot: a qualitative assessment of noise}\label{supsec:proplot}
Visually, the power spectrum in Fig. \ref{fig:fig1} appears to be flat (or white) between a few Hz to 1024 Hz, except for the bins surrounding the QPO feature at 224 Hz. Formally, a test for flatness or ``whiteness'' of a power spectrum is a test for whether the power spectral values are $\chi^2$ distributed \cite{vanderklis}. First, we assess this qualitatively by extracting a so-called probability plot. It shows the theoretical quantiles of an assumed distribution ($\chi^2$ with 2$\times$105$\times$2048 degrees of freedom (dof) scaled by a factor of 1/(105$\times$2048) in the present case) against the ordered sample values, i.e., observed noise powers. This particular $\chi^2$ distribution was used to compare because the PDS in the left panel of Fig. \ref{fig:fig1} was obtained by averaging 105 individual PDS followed by further averaging in frequency by a factor of 2048. If the observed data fall on a straight line on the probability plot then that indicates that they are consistent with the theoretical distribution that they are compared against. To remove the bias from the QPO we removed bins whose frequencies fall between 200 and 250 Hz. Also because we are really only interested in the nature of the noise continuum in the vicinity of 224 Hz we removed bins with frequencies above 600 Hz. Because there are many bins between 600 and 1024 Hz they could, in principle, skew the results. It is evident from Fig. \ref{fig:probplot} that the data points on the probability plot roughly follow a straight line (red line) and thus appear ``qualitatively'' consistent with a $\chi^2$ distribution with 2$\times$105$\times$2048 dof scaled by a factor of 1/(105$\times$2048). For completeness, we repeated this by considering all bins between 1/256 Hz and 1024 Hz and the result is the same.


\subsubsection{PDS continuum is consistent with White noise }\label{supsec:kstest}
To investigate the nature of the PDS continuum quantitatively, we performed the Kolmogorov-Smirnov (K-S) and the Anderson-Darling goodness-of-fit tests under the null hypothesis that the PDS continuum is white, i.e., the power values between 1/256 and 600 Hz, except for bins between 200-250 Hz, are $\chi^2$ distributed distributed with 2$\times$105$\times$2048 dof scaled by a factor of 1/(105$\times$2048). The basic idea with these statistics is that they measure the maximum deviation between the Empirical Distribution Function (EDF) of the data and that of a comparison distribution. Therefore, the better the distribution fits the data, the smaller these statistics will be. 

Before we evaluated the K-S and Anderson-Darling test statistics we computed the EDF and the probability density function (PDF) of the continuum noise powers. These are shown in the top two panels of Figure \ref{fig:kstest}) along with the expected EDF and PDF curves for a $\chi^2$ distribution (solid red). Similar to the probability plot, the data appear to track the expected $\chi^2$ distribution quite well.

We computed the K-S statistic using the EDF. To evaluate whether this value can be used to reject or not reject the null hypothesis, we calculated the distribution of K-S statistic values of EDFs drawn from the expected distribution: $\chi^2$ with 2$\times$105$\times$2048 dof scaled by a factor of 1/(105$\times$2048). We compute this K-S statistic distribution as follows. 
\begin{enumerate}
    \item First, we randomly draw 57 values from a $\chi^2$ with 2$\times$105$\times$2048 dof. Here, 57 refers to the total number of power spectral continuum values between 100 and 600 Hz and excluding those with frequencies between 200 and 250 Hz.
    \item Then we evaluate the EDF of this random sample of 57 and scale it by 1/(105$\times$2048)
    \item Finally, we compute its K-S statistic value.
\end{enumerate}
The above steps are repeated 10,000 times to get a distribution of the K-S test statistic for a $\chi^2$ distribution with 2$\times$105$\times$2048 dof for a given sample size of 57. This is shown as a blue histogram in the bottom left panel of Fig. \ref{fig:kstest}. AT2018cow's observed K-S test statistic (dashed vertical red line), which is a measure of maximum deviation between the observed EDF and the theoretical Cumulative Distribution Function (CDF), is lower than the typical value (solid magenta vertical line). This indicates that the null hypothesis cannot be rejected even at the 90\% confidence level and suggests that noise powers in the PDS continuum are very much consistent with the expected $\chi^2$ distribution, i.e., {\it NICER}'s PDS continuum is consistent with being white between 1/256 and 1024 Hz.

To ensure the above conclusion is not dependent on the choice of the statistic used we also computed the Anderson-Darling statistic. Similar to above, we computed its distribution using bootstrap simulations (see the bottom right panel of Fig. \ref{fig:kstest}). Again, it is clear that the statistic computed from the observed PDS of AT2018cow (vertical dashed red line) is consistent with the expected $\chi^2$ distribution. We repeated all the above tests by considering all the bins between 1/256 to the Nyquist frequency of 1024 Hz (excluding those between 200-250 Hz) and the results remained the same. We also varied the frequency upper limit from 400 Hz to 800 Hz and they all yield the same result.


\subsubsection{Modeling the PDS}\label{supsec:pdsmodel}
Next, we carried out a straightforward method to test for flatness of the PDS continuum. We fit the continuum, i.e., excluding the bins between 200 and 250 Hz, with two different models: a constant (white noise) and a constant + power-law (white + red noise). The model consisting of just a constant gave a $\chi^2$ of 129.9 with 121 dof while the constant+power-law yields a $\chi^2$ of 125.1 with 119 dof. We repeated this exercise with a PDS frequency resolution of 1/256 Hz, i.e., lowest frequency of 1/256 Hz. This also did not yield in significant $\chi^2$ improvement. This argues that a power-law component is not formally required by the data. 

Then we modelled the entire PDS (including the QPO bins) with a constant and a constant plus a Lorentzian to account for the QPO feature. While the former yielded a $\chi^2$ of 153.6 with 127 dof, adding a Lorentzian improved the $\chi^2$ by 23.9, i.e., resulted in a $\chi^2$ of 129.8 with 124 dof. This measurement forms the basis for our Monte Carlo simulations described in the following sections. The best-fit QPO centroid is 224.4$\pm$1.0 Hz respectively. 

The ratio of the sum of all the power values within the QPO width to error on that sum gives a quick estimate of the signal-to-noise of the QPO. For the 224 Hz feature this value is 0.0295/0.0061 = 4.8. Combined with the mean source count rate of 2.62$\pm$0.01 counts/sec, these values can also be used to estimate the fractional root-mean-squared (rms) amplitude of the QPO to be 100$\times\sqrt{0.0295*8/2.62}$ = 30$\pm$3\% (8 Hz is the frequency resolution here).


\subsubsection{Confirming white noise at lower (mHz) frequencies using longer/continuous light curves from {\it XMM-Newton}}\label{supsec:xmmpds}  
To ensure that AT2018cow's soft X-ray PDS is consistent with white noise at lower frequencies, i.e., 10$^{-3}$-10$^{-2}$ Hz we used a $\sim$ 30 ks {\it XMM-Newton} exposure that coincided with {\it NICER} monitoring. {\it NICER} data are not ideal for sampling at these low frequencies because of short exposures. We extracted an average 0.25-2.5 keV {\it XMM-Newton} PDS to find that there is no evidence for red noise down to in the frequency range of $\sim$10$^{-3}$ - a few Hz. The EPIC-pn PDS with a Nyquist frequency of 6.8 Hz is shown in Fig. \ref{fig:xmmpds}. 

Also, if there is strong red noise at low frequencies (or long timescales) that can manifest as variability on faster timescales. This effect is known as red noise leakage \cite{rednoiseleak}. In the present context, this means that if there is strong red noise below 1/256 Hz, that can, in principle, leak into the 100s of Hz range. The PDS in Fig. \ref{fig:xmmpds} also shows that the red noise leakage affect is negligible. 

\begin{figure}[ht]
\begin{center}
\includegraphics[width=6.5in, angle=0]{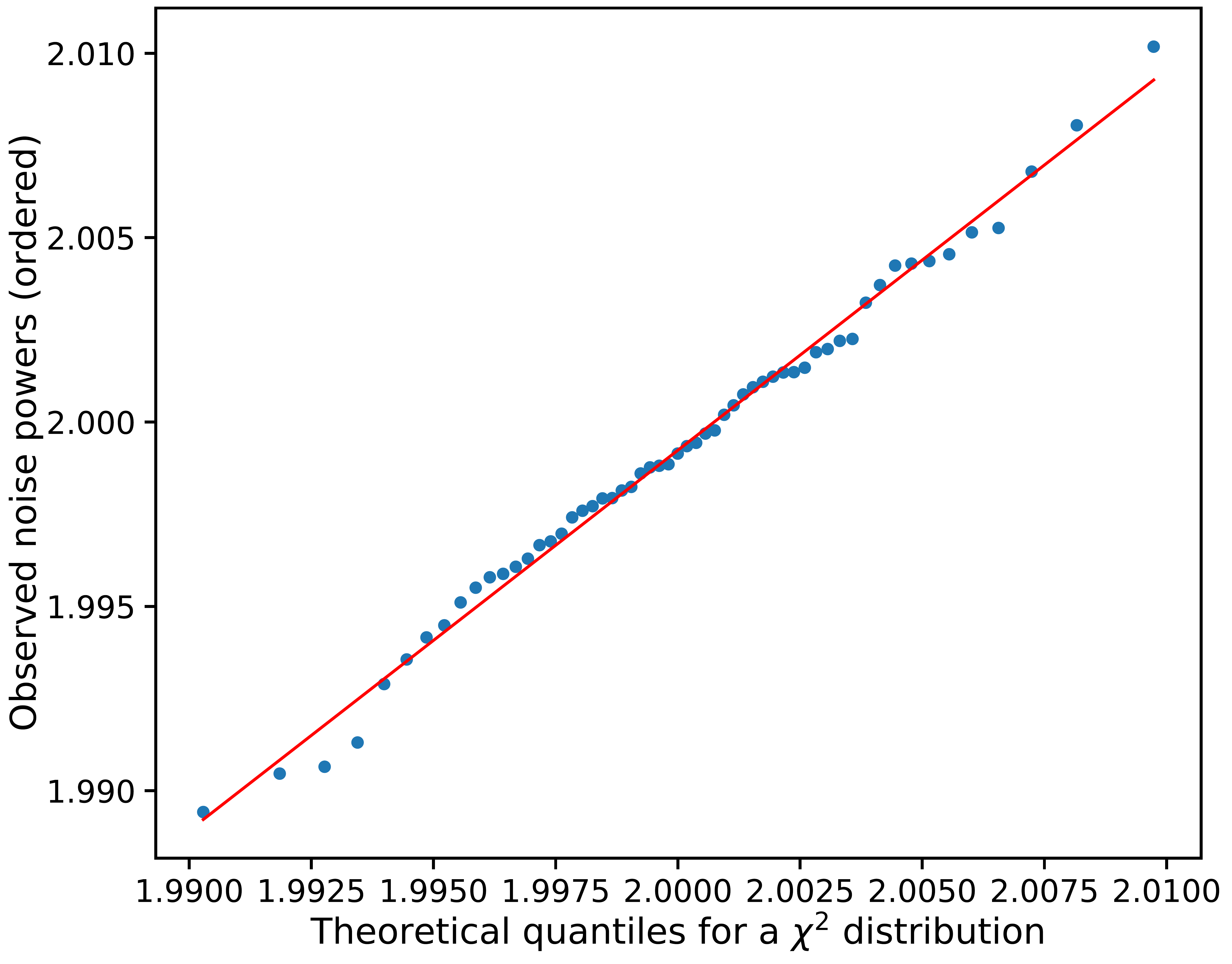}
\end{center}
\caption{{\textbf{ Probability plot to visually assess whether the power values in the PDS continuum are $\chi^2$ distributed.}} This is a {\it qualitative} assessment tool that tells us that if the data points lie on a straight line, as they do, they are appear consistent with the theorized model, which in the present case is a $\chi^2$ distribution with 2$\times$105$\times$2048 degrees of freedom (see section \ref{supsec:proplot} for more details).} \label{fig:probplot}
\end{figure}

\begin{figure}[ht]
\begin{center}
\includegraphics[width=6.25in, angle=0]{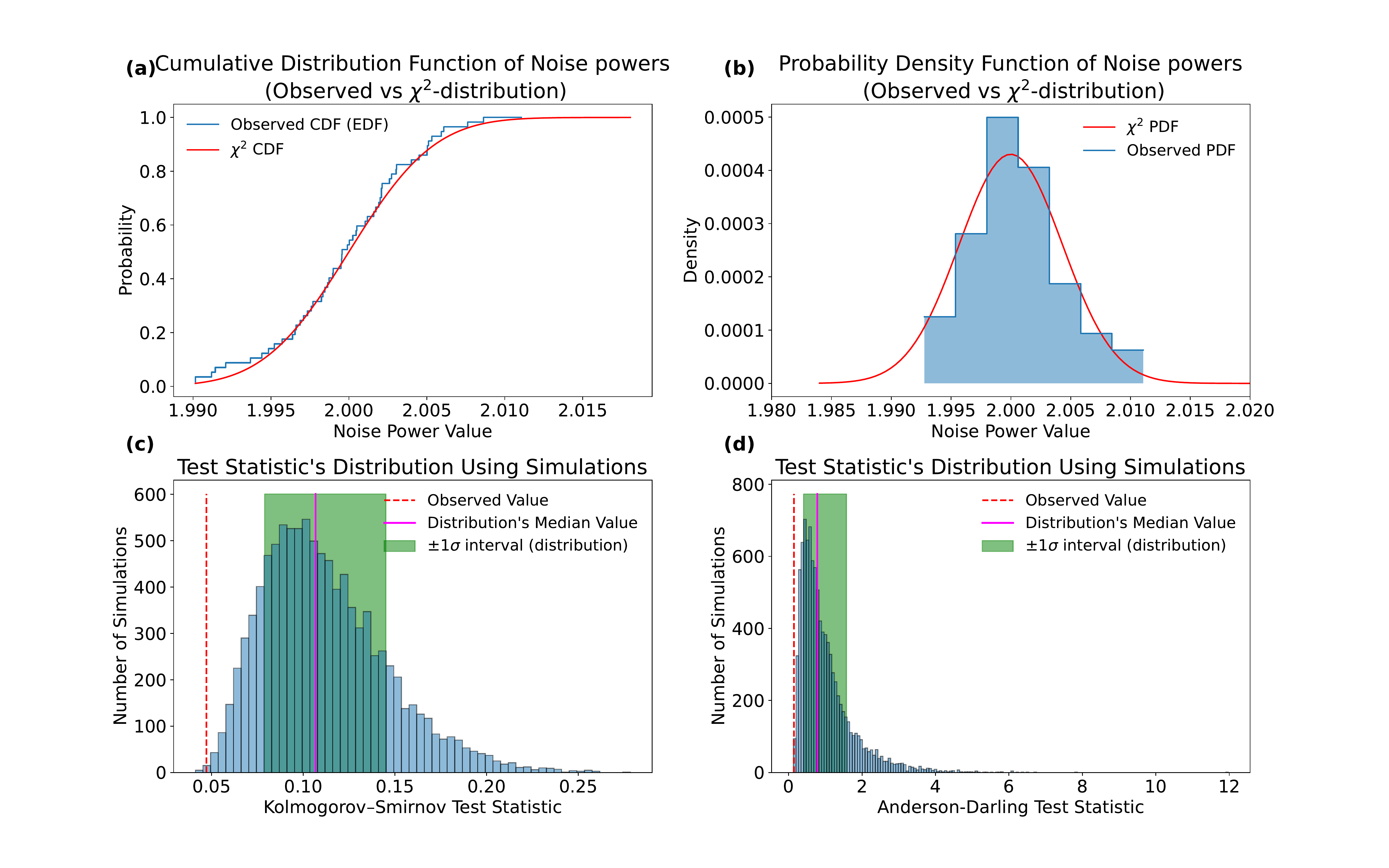}
\end{center}
\caption{{\bf White noise tests for soft X-ray PDS of AT2018cow}.{\it Top-left:} Comparing the empirical distribution function (EDF) of the values in the PDS continuum (blue histogram) with the cumulative distribution function of a $\chi^2$ distribution with 2$\times$105$\times$2048 degrees of freedom.  {\it Top-right:} Comparison of the observed probability distribution function (PDF) with the expected PDF of a $\chi^2$ distribution with 2$\times$105$\times$2048 dof. Both the observed PDF and EDF track expected curves quite well. {\it Bottom-left:} The distribution of the K-S statistic derived from EDFs sampled from a $\chi^2$ distribution with 2$\times$105$\times$2048 dof. {\it Bottom-right:} Same as {\it bottom-left} but using an Anderson-Darling test statistic. Both the test statistic values are consistent with a $\chi^2$ distribution and suggest that the PDS continuum is white (see section \ref{supsec:kstest} for more details).}\label{fig:kstest}
\end{figure}

\begin{figure}[ht]
\begin{center}
\includegraphics[width=0.85\textwidth, angle=0]{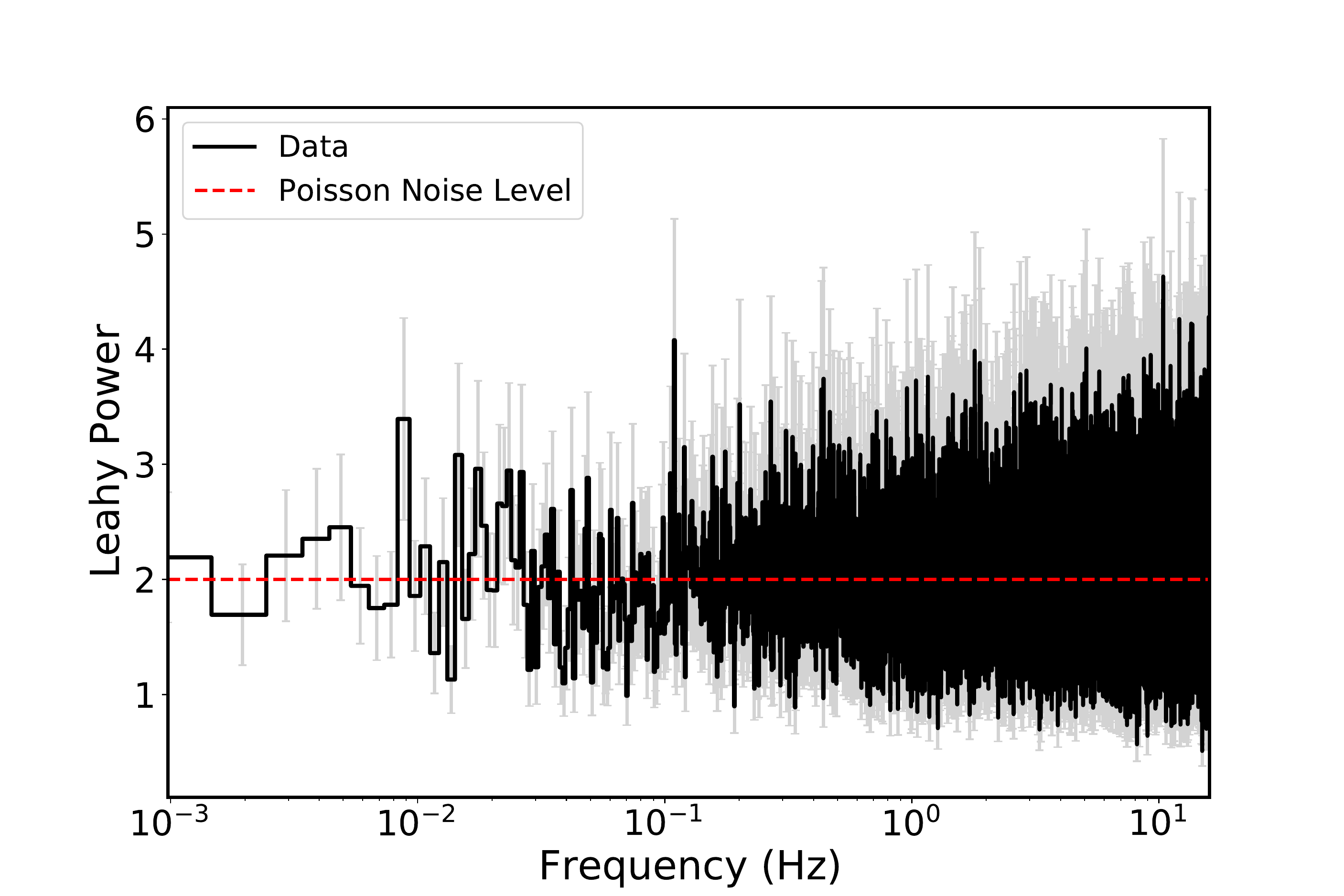}
\end{center}
\caption{{\bf {\it XMM-Newton}/EPIC-pn soft X-ray (0.25-2.5 keV) PDS of AT2018cow to assess noise continuum at low frequencies}. The PDS was derived by averaging 15 1024 s light curve segments sampled at 1/16 Hz. The frequency resolution is 1/2048 Hz. It is evident that even at frequencies as low as 1/2048 Hz there is no evidence for red noise. The 224 Hz QPO in {\it NICER} data is outside of this band pass, i.e., beyond EPIC-pn's Nyquist frequency.}\label{fig:xmmpds}
\end{figure}

Based on all the above described tests and Figs. \ref{fig:probplot}, \ref{fig:kstest} and \ref{fig:xmmpds} we conclude that over the frequency range of 1/256 to 1024 Hz the effect of red noise is negligible and that the PDS is consistent with being white.  


\subsubsection{Monte Carlo Simulations to Estimate Global Statistical Significance}\label{supsec:mcsims}
When the potential signal you are trying to test for is broad and distributed over several frequency bins, the standard approach of estimating significance based on just the highest bin will not suffice. As it will not include the contribution from all the QPO bins it will fail to capture the true significance estimate. Therefore, we devise an approach that can account for multiple frequency bins. The steps are as follows:
\begin{enumerate}
    \item After establishing that AT2018cow's soft X-ray variability, on timescales of 1/256 Hz to 1024 Hz is white, i.e., frequency-independent, we generate a set of 105 256-s ``simulated'' light curves by simply randomly shuffling each observed light curve independently, i.e., time bins do not move from one segment to another. In practice (in {\tt Python}) this is done using {\tt numpy}'s {\tt random.shuffle} function.
    \item Then we extracted an ''simulated'' average PDS from these shuffled light curves and re-binned it to a frequency resolution of 8 Hz just like the real PDS in Fig. \ref{fig:fig1}. 
    \item We then searched for a QPO within this simulated average PDS at all frequencies, i.e., we fit a constant and a constant + a Lorentzian model with the centroid constrained to the ends of each frequency bin and the width allowed to be free, and computed an array of $\Delta\chi^2$ values. 
    \item Finally, we record the maximum $\Delta\chi^2$ value from the array of $\Delta\chi^2$ value from step 3 
\end{enumerate}

\begin{figure}[ht]
\begin{center}
\includegraphics[width=\textwidth, angle=0]{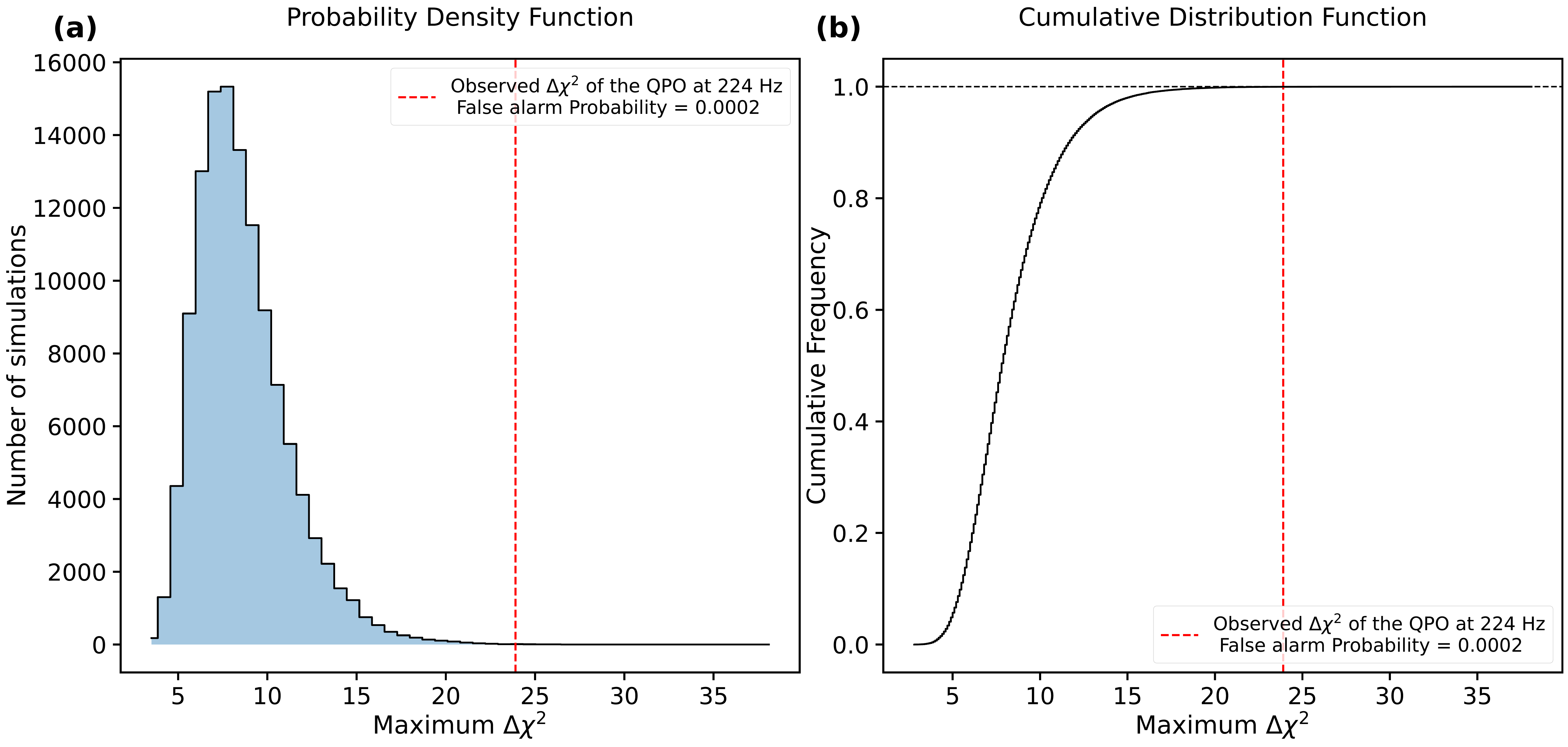}
\end{center}
\caption{{\bf The probability density function (PDF) (left) and the cumulative distribution function (CDF) (right) of the 10$^{5}$ simulated maximum $\Delta\chi^{2}$ values}. The vertical dashed red line marks the value of the $\Delta\chi^{2}$ of the observed 224 Hz QPO. The false alarm probability (1-CDF) vs $\Delta\chi^2$ is shown in the right panel of Fig. \ref{fig:fig1}. The false alarm probability of seeing a spurious signal as strong as the one seen in data at 224 Hz is 0.0002.}\label{fig:pdcdfs}
\end{figure}

The above steps were repeated 10$^{5}$ times to get an array of 10$^{5}$ maximum $\Delta\chi^2$ values. We used 8 cores on a laptop for these simulations which took a total of roughly 180 hours for 10$^{5}$ simulations. From these measurements we computed the probability to exceed a certain $\Delta\chi^2$ value, i.e., 1-CDF. This is shown in the right panel of Fig. \ref{fig:fig1}. For guidance, the 99.73\% and 99.99\% confidence levels are indicated.  Another way to realize the results of the simulations above is through the probability density function (PDF) and the CDF of the maximum $\Delta\chi^2$ values. These are shown in Fig. \ref{fig:pdcdfs}. The QPO found in {\it NICER} data is statistically significant at roughly 2$\times$10$^{-4}$ level which translates to 3.7$\sigma$ equivalent for a normal distribution.

 
\subsubsection{{\it XMM-Newton}/EPIC and {\it Swift}/XRT data cannot be used for finding/studying the 224 Hz QPO}
{\it XMM-Newton}/EPIC's pn and MOS data were taken in the so-called full-frame mode which provide a time resolution of 73.4 ms and 2.6 s, respectively. The majority of {\it Swift}/XRT data were taken in the Photon Counting (PC) mode which has a time resolution of about 2.5 s. Thus, these datasets have Nyquist frequencies that are significantly lower than the QPO's frequency of 224 Hz. As a consequence, EPIC and XRT data cannot be used to search for and study the 224 Hz QPO found in {\it NICER} data.


\subsection{Ruling out an instrumental and a non-astrophysical origin}\label{supsec:instru}
All {\it NICER} events have two types of pulse height amplitude (PHA) data: the ``slow'' PHA derived from the slow chain electronics and the ``fast'' PHA (PHA\_FAST) derived by the fast chain electronics. The standard {\it NICER} calibration scripts convert this information into pulse invariant (PI) and PI\_FAST, respectively. For events with energies $\lesssim$ 600 eV the fast-chain is not triggered and thus PI\_FAST is undefined. Although {\it NICER}’s XTI is a non-imaging instrument, there is a way to separate the background (particles, cosmic X-rays, and low-energy/optical light loading events) and the source events using the values of PI and PI\_FAST. The X-ray events from an on-axis astrophysical source, high-energy particles, light loading, and the cosmic X-ray background occupy a separate region of the PI vs PI\_RATIO (PI/PI\_FAST) plot\footnote{\url{https://heasarc.gsfc.nasa.gov/docs/nicer/mission_guide/}}. 

To ensure the detected signal does not originate from any of these three types of background, we extracted an average PDS of all three backgrounds using their corresponding events within the GTIs. There was no evidence of a variability enhancement around 224 Hz (or elsewhere) in any of these three backgrounds (see Fig. \ref{fig:noisepds}). We describe each of these analyses in a bit more detail below.


\subsubsection{The signal is not present in the particle background}
High energy particles from space can interact with the silicon material in {\it NICER}/XTI’s SDDs to produce charge and mimic X-ray events. These particle events are registered as enhancements in the so-called overshoot rate for each FPM. Also, because particles have energies much higher than {\it NICER}’s nominal bandpass of 0.25-12 keV their incidence is also evident in the 13-15 keV count rate. Thus, if this QPO signal were from background particle events it would also be present in the overshoot and the 13-15 keV data. 

\begin{figure}[ht]
\begin{center}
\includegraphics[width=\textwidth, angle=0]{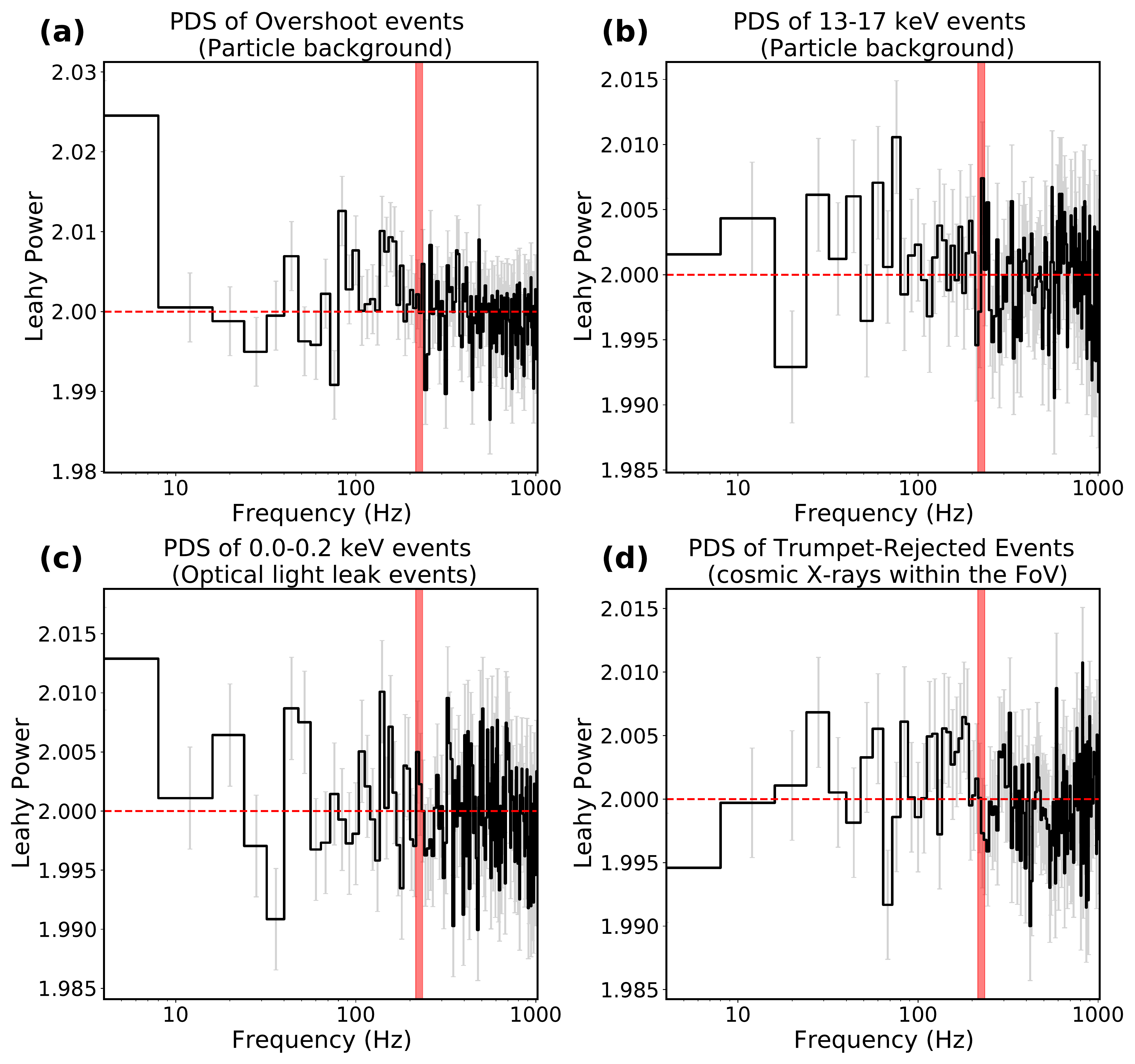}
\end{center}
\caption{{\bf {\it NICER} PDS of various types of noise.} {\bf (a) and (b) show the PDS of the overshoots and the 13-15 keV events, respectively.} They both track the particle background. {\bf (c) The average PDS of 0.0-0.2 keV events.} These are the optical light leak events. {\bf (d) The average PDS derived from trumpet-rejected events.} These are the off-axis events that track the cosmic X-ray and other sources in the FoV. In all cases, the frequency resolution, the number of spectra averaged and the frequency rebinning is same as the left panel of Fig. \ref{fig:fig1}. The red shaded rectangle shows the location of the 224 Hz QPO. Clearly, there is no statistically significant evidence for a 224 Hz QPO in any of these noise power spectra. The dashed horizontal line indicates the expected Poisson noise level of 2. These plots suggest that the 224 Hz QPO in Fig. \ref{fig:fig1} does not originate from any of these background events.}\label{fig:noisepds}
\end{figure}

We extracted the overshoot only event lists using the {\it NICER} data reduction pipeline described above but with a modified event flag EVENT\_FLAG={\tt bxxx01x}. This particular choice selects only the overshoot events. Then, similar to our analysis on source events, we applied barycenter correction on these unfiltered (but calibrated) events. Because overshoots do not contain a PI value they cannot be energy calibrated like real X-ray events. We then extracted the average PDS using only the events within the standard GTIs (see the top-left panel of Fig. \ref{fig:noisepds}). There is no evidence for an excess variability anywhere in the PDS.

High-energy events with energies in between 13 and 15 keV that includes the {\it trumpet} and all PI\_RATIOs were extracted using the standard {\it NICER} tools. Then they were barycenter-corrected using the {\tt barycorr} tool similar to the source events. The average PDS from these events is shown in the top-right panel of Fig. \ref{fig:noisepds} and does not show any obvious QPO like features anywhere in the considered frequency range.

The average count rate of the overshoots and the 13-17 keV were 18.7 and 0.18 counts/sec, respectively. The former value is much higher than AT2018cow's mean soft X-ray count rate of 2.62 counts/s. Therefore, if the 224 Hz QPO were from background particles it would have shown up in the top-left panel of Fig. \ref{fig:noisepds} (sensitivity towards a QPO increases linearly with count rate \cite{vanderklis}). Thus, the top two panels of Fig. \ref{fig:noisepds} allows us to rule out a particle background origin for the 224 Hz QPO.


\subsubsection{The signal is not due to optical light leak events}
Optical light, either directly from the Sun or bright Earth at low angles, or from reflections off ISS surfaces can make its way into the FPMs. This manifests as a noise peak in each FPM whose energy spectrum peaks at an energy below 0.25 keV. However, when this light loading, which is time variable, is strong, the tail end of the noise distribution can leak into higher energies ($>$0.25 keV) and contaminate the source events. This effect is especially important for faint targets like AT2018cow. The event filtering described in section \ref{supsec:nicerdatareduc} should already remove epochs of high light loading. To further ensure that the QPO signal is not a result of variability of ``light leak'' events, if any, we extracted a 0.0-0.2 keV PDS. Similar to the inband PDS we first extracted the 0.0-0.2 keV events using the standard {\it NICER} tasks. We then barycenter-corrected them and then computed an average PDS using events within the standard GTIs. The resulting power spectrum shown in the bottom-left panel of Fig. \ref{fig:noisepds} allows us to rule out a light leak origin for this QPO.


\subsubsection{The signal is not present in cosmic X-ray background events within {\it NICER}'s Field of View (FoV)}
Although {\it NICER}/XTI is a non-imaging instrument, it's design with the slow and the fast chain electronics allows us to separate on-axis events from the off-axis ones. While the on-axis events make a ``Trumpet''-like cluster in the PI vs PI\_RATIO plot, the off-axis events, i.e., those from cosmic X-rays and other point sources in the FoV, fall above the ``Trumpet''. If the QPO signal is associated with AT2018cow (placed on-axis during each observation) then it should not be present in the trumpet-rejected events. The inband PDS shown in Fig. \ref{fig:fig1} already excludes trumpet-reject events. However, to be absolutely sure, we extract an average PDS of the trumpet-rejected events within the standard GTIs. We extracted the trumpet-rejected events by first extracting the standard cleaned event lists with {\tt trumpfilt=NO}. These were barycenter-corrected and then screened to only include events between 0.25 and 10 keV that fell above the trumpet edge defined by the curve PI\_RATIO = 1.1 + 120/PI. An average PDS was extracted in same manner as the inband PDS. This is shown in the bottom-right panel of Fig. \ref{fig:noisepds} and it is evident that a 224 Hz QPO is not present in events that describe the cosmic X-ray background and any contaminating sources in the field of view.


\subsubsection{The signal is present in all Measurement and Power Units (MPUs)}
\begin{figure}[ht]
\begin{center}
\includegraphics[width=\textwidth, angle=0]{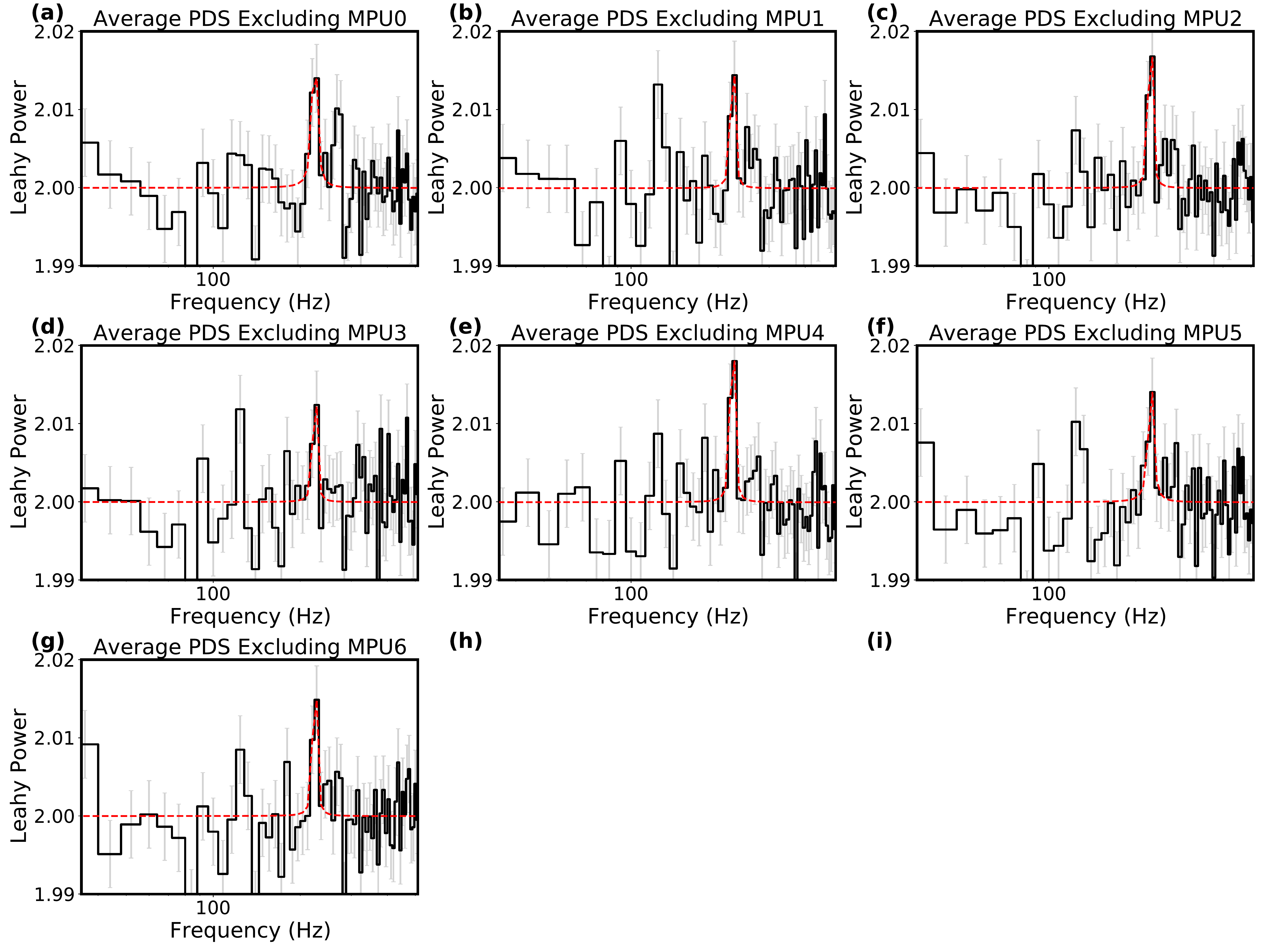}
\end{center}
\caption{{\bf {\it NICER} power spectra with one MPU removed at a time.} The QPO properties, i.e., fractional rms strength, centroid and width are all same (within errorbars) across all the above PDS. This demonstrates that the QPO signal is uniformly distributed across all the MPUs and points towards an non-instrumental origin.}\label{fig:mpupds}
\end{figure}

If this signal is astrophysical in origin then it must be uniformly distributed across all the 52 active FPMs which are controlled by 7 MPUs$\footnote{\url{https://heasarc.gsfc.nasa.gov/docs/nicer/mission_guide/}}$. We checked to ensure that is the case by extracting several average PDS, but with data from one MPU removed at a time. The resulting 7 PDS with the ID of the removed MPU at the top are shown in Fig. \ref{fig:mpupds}. The QPO is evident in all the 7 PDS with a fractional rms amplitude value consistent with each other. This demonstrates that the QPO events are uniformly distributed across all MPUs and not just limited to any single unit.


\subsubsection{The signal is not due to ``GPS'' noise}\label{supsec:gps}
\begin{figure}[ht]
\begin{center}
\includegraphics[width=0.6\textwidth, angle=0]{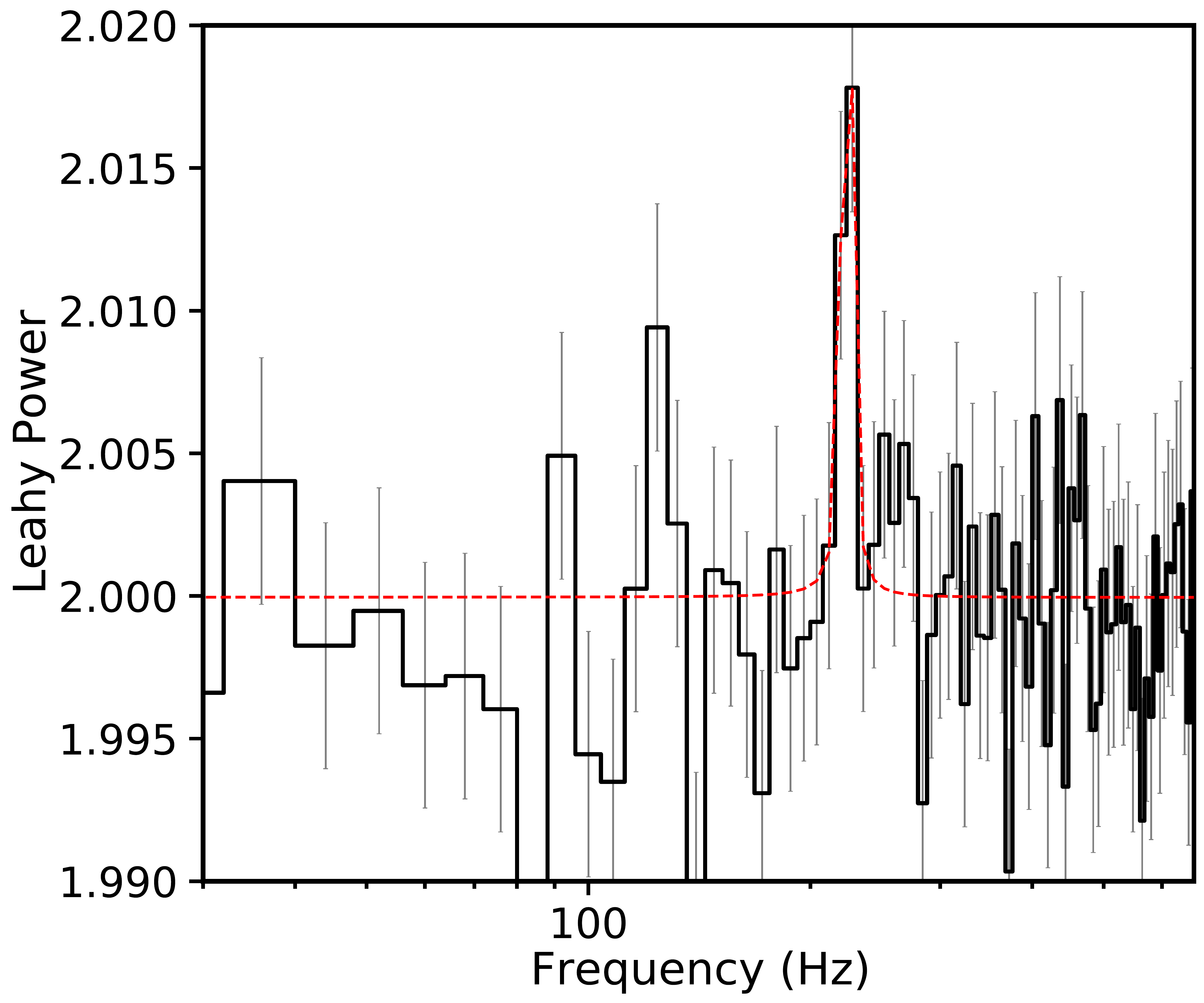}
\end{center}
\caption{{\bf Same power spectra as the left panel of Fig. \ref{fig:fig1} but with plausible GPS noise events removed.} This PDS is indistinguishable from Fig. \ref{fig:fig1} and demonstrates that the GPS noise, if any, is not the origin of the 224 Hz QPO (see section \ref{supsec:gps} for more details)}\label{fig:gpspds}
\end{figure}

\begin{figure}[ht]
\begin{center}
\includegraphics[width=\textwidth, angle=0]{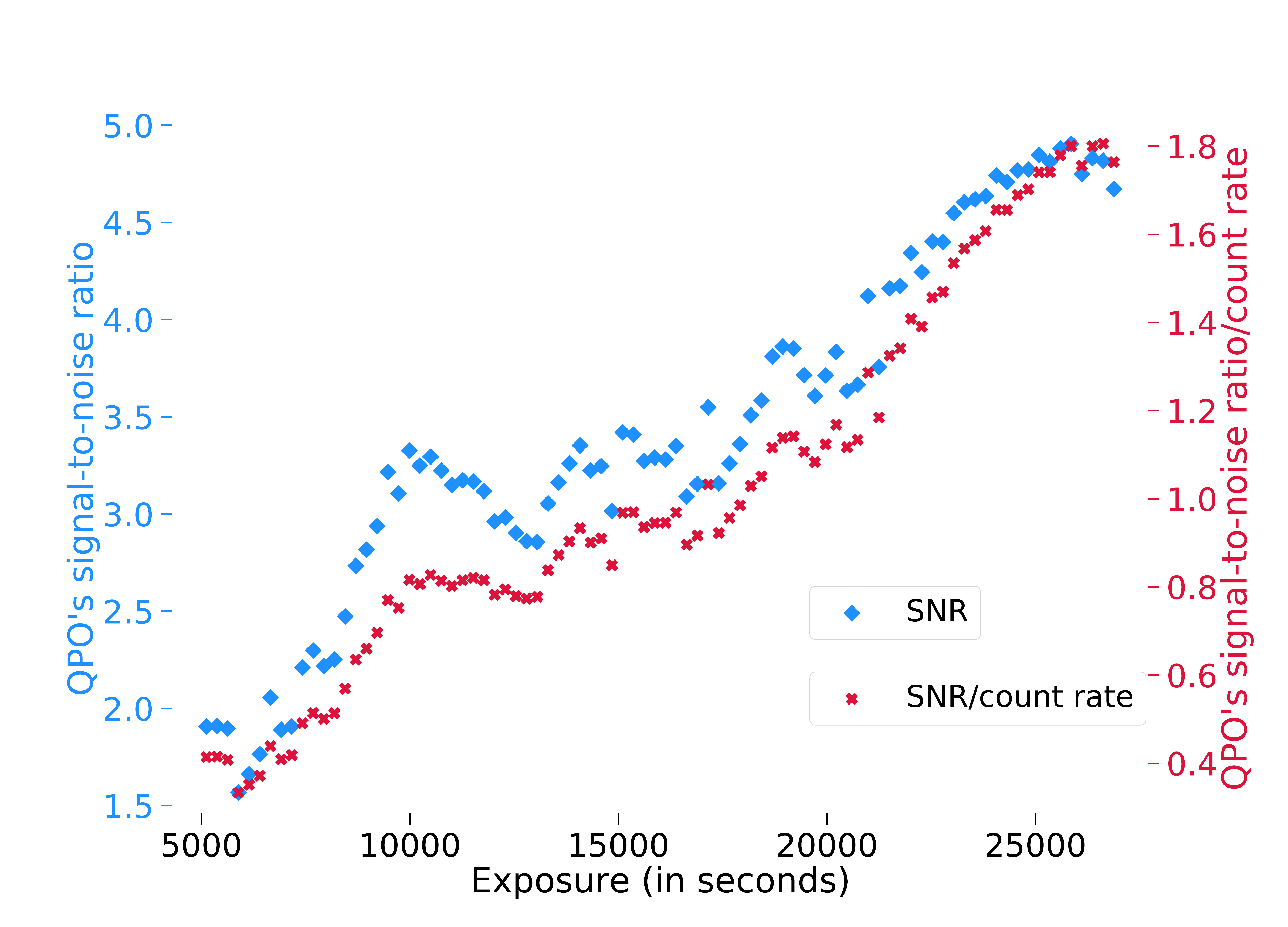}
\end{center}
\caption{{\bf QPO's signal-to-noise ratio (blue diamonds) and signal-to-noise over average count rate (red crosses) vs accumulated exposure time.} It is evident that the QPO's signal-to-noise gradually increases with increasing exposure. This suggests that the signal is persistent throughout the {\it NICER} monitoring period and it does not originate from any single exposure. The steepening of the mean slope around 20000 s corresponds to day 17. Note that the data points are not independent (see sec. \ref{supsec:qpoexpos} for more details). See also supplement movie S1. Data available in supplementary files.}\label{fig:deltachi}
\end{figure}

One other plausible instrument-related source of origin of this signal is the co-called ``Global Position System (GPS) noise’’. During the ground testing of {\it NICER}’s MPUs anomalous cross-talk signals between the FPM input line and the GPS pulse-per-second (PPS) line were observed. This can result in certain FPMs registering pseudo events (non-cosmic) immediately following a GPS PPS tick, usually within the first 4 ms after the tick. In principle, this could produce a 1 second QPO with roughly a 4 ms QPO width, although it has never been reported in any of the analysis of several {\it NICER} targets thus far. This is obviously at a different frequency compared to the 224 Hz QPO from observations of AT2018cow. Nevertheless, we rule out GPS noise as the source of AT2018cow’s QPO by first removing all events with time stamps occurring within 10 ms of an integer second of the original non-barycenter corrected data. Then, we barycenter corrected the remaining events and extracted an average PDS in the exact same manner as the PDS in Fig. \ref{fig:fig1}. The resulting PDS shown in Fig. \ref{fig:gpspds} is virtually indistinguishable from the left panel of Fig. \ref{fig:fig1}. This is not too surprising as the above only excludes a small fraction of all events. Fig. \ref{fig:gpspds} shows that the signal is still present after removing the plausible GPS noise events and rules out GPS noise as the origin of this QPO.


\subsubsection{AT2018cow dominates the X-ray emission in {\it NICER}/XTI's FoV}\label{supsec:image}
{\it NICER}/XTI is a non-imaging detector with a field of view of roughly 30 arcmin$^{2}$. Thus it is plausible that a contaminating point source other than AT2018cow could have produced this QPO signal. To investigate this possibility, we extracted an image by stacking the entire Neil Gehrels {\it Swift}/XRT archival data of AT2018cow (see the right panel of Fig. \ref{fig:fig3}). It is clear from these images that AT2018cow is the brightest point source within XTI's FoV. The contribution from the other point sources is negligible. 

Furthermore, we can directly compare {\it NICER}/XTI light curve of AT2018cow's FoV with resolved XRT light curve. The variability features in XRT light curve are also evident in XTI data (see Fig. \ref{fig:fig2}). This independently implies that AT2018cow dominates the X-ray emission in XTI's FoV.

As a final check, we also investigated a late time {\it XMM-Newton} image of AT2018cow's FoV to rule out contamination by a point source within XRT or EPIC instrument's point spread functions. The left panel of Fig. \ref{fig:fig3} show MOS1 image of AT2018cow's FoV at late times. We used MOS1 because it offers the best pixel size of 4.1''. It is evident that long after AT2018cow's optical decay, there is no X-ray emission present at its location and rules out a contaminating source very close to the position of AT2018cow. 

The combination of the above three analyses affirms that the majority of emission detected by XTI between MJD 58290.87 and 58349.11 and thus the 224 Hz signal originates from AT2018cow.

\subsubsection{A similar signal is NOT present in any AGN monitored by {\it NICER} during the same epoch}
\begin{figure}[ht]
\begin{center}
\includegraphics[width=0.4\textwidth, angle=0]{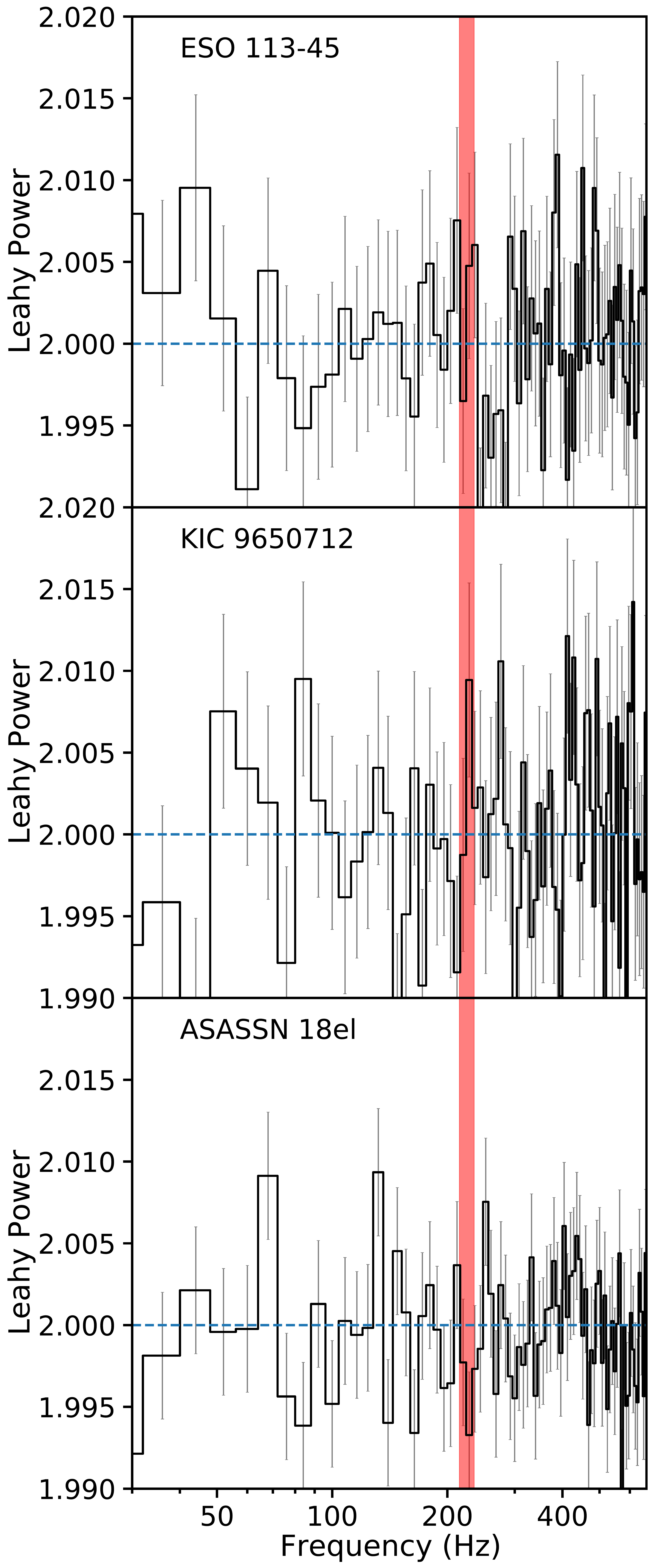}
\end{center}
\caption{{\bf Average PDS of 3 AGN monitored by {\it NICER} during the same 2 month period as AT2018cow.} The AGN names are shown on each panel. The mean count rates (exposures) (from top to bottom) are 30.5 (15.6 ks), 3.9 (14.3 ks) and 1.2 (33.3 ks)  counts/s, respectively. These PDS were extracted exactly the same way as the average PDS of  AT2018cow in Fig. \ref{fig:fig1}. The location of AT2018cow’s QPO is indicated by the red shaded area. There is no evidence for a 224 Hz QPO in any of these power spectra. This provides further support that the 224 Hz QPO in Fig. \ref{fig:fig1} is intrinsic to AT2018cow's data.}\label{fig:agnpds}
\end{figure}

AGN host supermassive black holes ($\gtrsim$10$^{5}$M$_{\odot}$) and hence causality argument suggests that they should not show any variability on timescales of milliseconds, i.e., hundreds of Hz. Therefore, to be absolutely sure that the 224 Hz QPO signal is unique to AT2018cow data we also extracted average PDS of 3 active galactic nuclei that {\it NICER} monitored during the same epoch, i.e., only using their observations between MJD 58290.87 and 58349.11. These resulting average PDS are shown in Fig. \ref{fig:agnpds} and a 224 Hz feature is not seen in any of them. 

\subsection{QPO is highly persistent}\label{supsec:qpoexpos}
The fact that the QPO is detected in the average PDS extracted from data accumulated over a temporal baseline of $\approx$60 days suggests that the QPO is likely present for a significant fraction of the X-ray monitoring. To further test this we estimated the strength (signal-to-noise ratio) of the QPO as a function of the accumulated exposure. We started with the first 5 ks of exposure, extracted an average PDS, and fit the 224 Hz feature with a Lorentzian. From thereon, for every additional 256-s exposure, we repeated this process of extracting an average PDS, followed by modelling the 224 Hz feature with a Lorentzian. At every point, the QPO strength was calculated as the ratio of the sum of the powers within the width of the best-fit QPO and its errorbar. A normalized QPO strength was also estimated by dividing the QPO strength with the average count rate which itself evolves with the accumulated exposure. It is evident from Fig. \ref{fig:deltachi} that the overall strength of the QPO gradually increases with increasing exposure. This demonstrates that:

\begin{enumerate}
    \item The QPO was persistent for a substantial fraction of the 60 d monitoring program, i.e., for one billion cycles ($\sim$60days/4.4ms). 
    \item The QPO signal does not originate from any single exposure of the {\it NICER} monitoring.
\end{enumerate}

The same result can be visualized in the form of a movie showing the evolution of the average PDS as a function of increasing exposure. This can be found as a supplementary file (Movie S1). A gif version is also available. 


\section{Upper limits on fractional rms of harmonics}\label{supsec:harmonics}
\begin{figure}[ht]
\begin{center}
\includegraphics[width=\textwidth, angle=0]{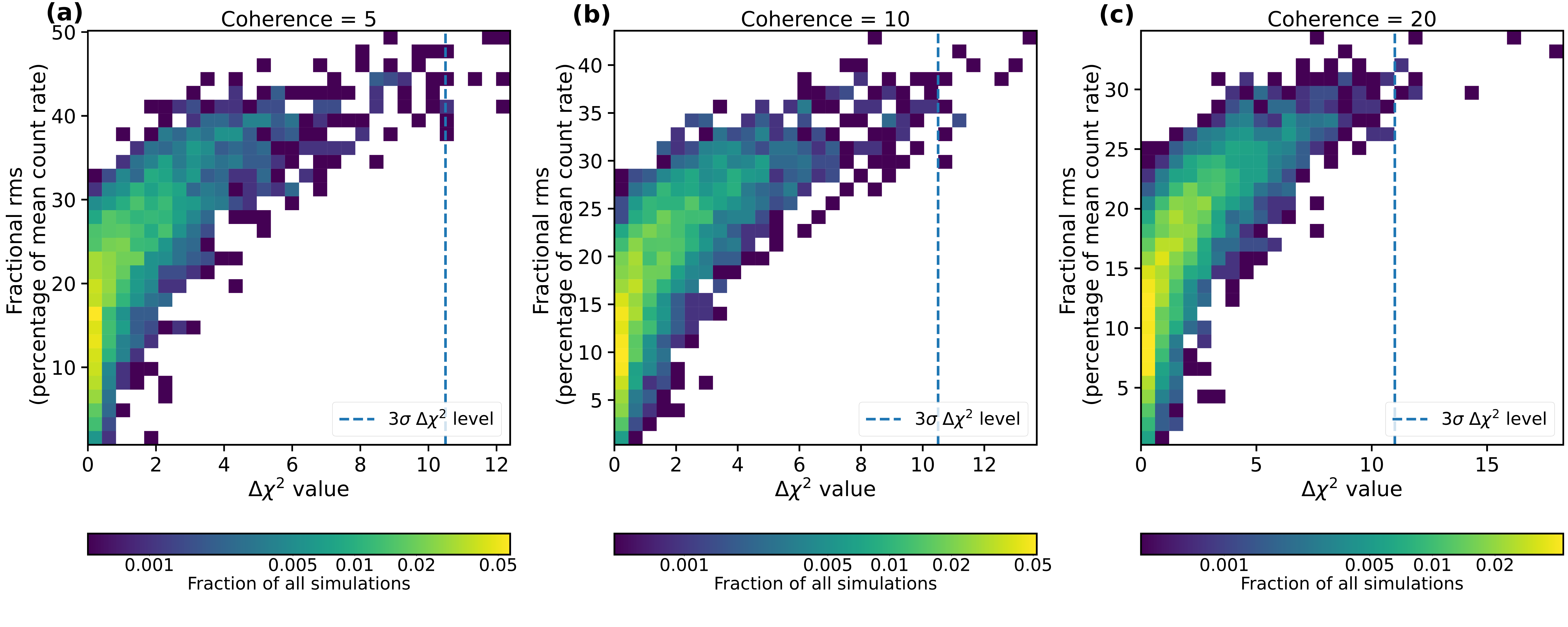}
\end{center}
\caption{{\bf 2-D histograms of $\Delta\chi^{2}$ improvement (constant vs constant + Lorenztian) vs corresponding fractional rms of a QPO-like feature at half the observed frequency, i.e., between 104-120 Hz.} These were derived using simulations (See sec. \ref{supsec:harmonics} for more details). The dashed vertical line in each panel corresponds to the 3$\sigma$ (99.73\%) level. The upper limit on fractional rms corresponds to the intersection of the histogram with the 3$\sigma$ vertical line (see Table. \ref{tab:rmstab}). The three panels correspond to three different coherence values (centroid frequency/width) of a QPO-like feature between 104 and 120 Hz.}\label{fig:rmsupp}
\end{figure}

We estimate the 3$\sigma$ upper limits on the fractional rms of potential QPOs near the  harmonic frequencies of the 224 Hz QPO using a similar Monte Carlo methodology described in sec. \ref{supsec:mcsims}. The procedure for estimating upper limit on fractional rms near $\frac{1}{2}\times$(224$\pm$16) Hz is described below but the same methodology was employed for estimating QPO rms upper limits near  $\frac{3}{2}\times$(224$\pm$16) Hz and $2\times$(224$\pm$16) Hz.
\begin{enumerate}
    \item First, we simulate a noise power spectrum exactly as described in steps 1-2 of sec. \ref{supsec:mcsims}.
    \item Then, we fit a Lorentzian with centroid constrained between 104 and 120 Hz (224$\pm$16/2) and width fixed to a value corresponding to a coherence value of 5. The value of the fractional rms and the $\Delta\chi^{2}$ estimated from the best-fit Lorentzian parameters is recorded. 
    \item The above steps are repeated 5000 times to get two arrays: one for $\Delta\chi^{2}$ values and another for fractional rms estimates (see the left panel of Fig. \ref{fig:rmsupp}). We discard all values corresponding to simulations that result in a dip in the PDS near 110 Hz.
    \item Using the simulated $\Delta\chi^{2}$ values above we estimate the $\Delta\chi^{2}$ value corresponding to 99.73\% (3$\sigma$). 
    \item Finally, using the mapping between the $\Delta\chi^{2}$ vs fractional rms values (see left panel of Fig. \ref{fig:rmsupp}) we estimate the fractional rms corresponding to the value of $\Delta\chi^{2}$ which marks the 3$\sigma$ level.
\end{enumerate}
The same steps were repeated for estimating the rms upper limit with coherence values of 10 and 20. The 3$\sigma$ upper limits from the above analysis are tabulated in Table. \ref{tab:rmstab}.

\section{Implications for a magnetar scenario}\label{supsec:ns}
The bolometric light curve of AT2018cow reached a luminosity of $\sim$ 4$\times$10$^{44}$ erg s$^{-1}$ on time a timescale of a few days after the explosion, and the luminosity decayed as $ L_{\rm e} \propto t^{-\alpha} $, where $\alpha$ $\approx$ 2.5 at times, t $\gtrsim$ t$_{e}$ $\sim$ 10$^{3}$ - 10$^{5}$ s. The central engine responsible for powering the optical
and X-ray emission must supply a total energy E$_{\rm e}$ $\sim$ 10$^{50}$ - 10$^{52}$ erg over a characteristic timescale of t$_{\rm e}$. Degeneracy exists in these properties because we do not know how much of the kinetic energy of the ejecta is supplied by the initial explosion itself,
versus injected at later times by the engine.

Two models for the central engine include: (1) a stellar-mass
black hole of mass $\sim$ 10-30M$_{\odot}$ created by the failed explosion of
a very massive star, which is accreting fall-back material from the
outer layers of the extended progenitor at a highly super-Eddington
rate; (2) a magnetar with a rapid birth period P$_{\rm 0}$ $\sim$ ms and a strong
dipole magnetic field B $\sim$ 10$^{14}$-10$^{15}$ G. The magnetar may be
spinning down in isolation, or it may be accreting fall-back material
as in the black hole. The latter scenario may be supported by the
predicted decline rate of the engine luminosity  $ L_{\rm e} \propto t^{-2.38} $.

We now consider models for the origin of the QPO-like feature
at f$_{\rm QPO}$ = 224 Hz, assuming it is related to the spin period of a
magnetar, P$_{\rm QPO}$ = 1/ f$_{\rm QPO}$ $\approx$  4.44 ms or half of that value (P$_{\rm QPO}$ =
8.9 ms). Alternatively, the QPO could arise directly in the accretion
disk (of the black hole or neutron star), e.g. as in an X-ray binary QPO, a
possibility which is not addressed below.

\subsection{Spin of an isolated pulsar or magnetar}
Here we consider the possibility that the observed QPO at 224~Hz is the spin frequency of a pulsar or magnetar.  The traditional picture of pulsar spin evolution  (see, e.g., \cite{Shapiro83}, for a review) is that the pulsar's rotational energy
\begin{equation}
E_{\rm rot}(\nu)=\frac{I}{2}\left(2\pi\nu\right)^2\sim1\times10^{51}\mbox{ erg}\left(\frac{\nu}{224\mbox{ Hz}}\right)^2,
\end{equation}
where $I\sim10^{45}\mbox{ g cm$^{2}$}$ is the stellar moment of inertia, can supply energy at a rate
\begin{equation}
L_{\rm sd}=4\pi^2I\nu\dot{\nu}\sim3\times 10^{43}\mbox{ erg s$^{-1}$}\left(\frac{\nu}{224\mbox{ Hz}}\right)\left(\frac{\dot{\nu}}{3\times 10^{-6}\mbox{ Hz s$^{-1}$}}\right)
\end{equation}
to power radiation from a rotating magnetic dipole at a rate $L_{\rm mag}\propto B^2\nu^4$. Equating the two rates yields an estimate of the pulsar magnetic field
\begin{equation}
B\approx2\times 10^{13}\mbox{ G}\left(\frac{\nu}{224\mbox{ Hz}}\right)^{-3/2}\left(\frac{-\dot{\nu}}{3\times 10^{-6}\mbox{ Hz s$^{-1}$}}\right)^{1/2} \label{eq:magb}
\end{equation}
and evolution of spin frequency
\begin{equation}
\nu=\nu_0(1+t/t_{\rm sd})^{-1/2}, \label{eq:eq6}
\end{equation}
which then leads to
\begin{equation}
L_{\rm sd}=\frac{E_{\rm rot}(\nu_0)}{t_{\rm sd}}\frac{1}{(1+t/t_{\rm sd})^{2}}=\frac{3\times 10^{43}\mbox{ erg s$^{-1}$}}{(1+t/t_{\rm sd})^{2}}\left(\frac{\nu_0}{224\mbox{ Hz}}\right)\left(\frac{-\dot{\nu}_0}{3\times 10^{-6}\mbox{ Hz s$^{-1}$}}\right), \label{eq:edot}
\end{equation}
where the characteristic timescale for spin-down is
\begin{equation}
t_{\rm sd}=-\frac{\nu_0}{2\dot{\nu}_0}\approx430\mbox{ d}\left(\frac{\nu_0}{224\mbox{ Hz}}\right)\left(\frac{-\dot{\nu}_0}{3\times 10^{-6}\mbox{ Hz s$^{-1}$}}\right)^{-1}\!\!=430\mbox{ d}\left(\frac{\nu_0}{224\mbox{ Hz}}\right)^{-2}\left(\frac{B}{2\times10^{13}\mbox{ G}}\right)^{-2}
\end{equation}
and $\nu_0$ and $\dot{\nu}_0$ are the initial spin frequency and spin frequency time derivative, respectively.  One can see from equation~(\ref{eq:edot}) that the luminosity changes little at times $t\ll t_{\rm sd}$ and decreases as $L_{\rm sd}\propto t^{-2}$ when $t\gg t_{\rm sd}$.

The QPO in AT2018cow is observed to persist at 224~Hz for about 60 days
with a width of $<16\mbox{ Hz}$.  This constrains the frequency change to
\begin{equation}
|\dot{\nu}|<16\mbox{ Hz}/\mbox{60 d}=3\times 10^{-6}\mbox{ Hz s$^{-1}$} \label{eq:nudot}
\end{equation}
and magnetic field and spin-down timescale to $B<2\times10^{13}\mbox{ G}$ and $t_{\rm sd}>430\mbox{ d}$, respectively.  Even stricter constraints of
\begin{equation}
-\dot{\nu}<3\times 10^{-8}\mbox{ Hz s$^{-1}$}\left(\frac{L}{3\times10^{41}\mbox{ erg s$^{-1}$}}\right)
\end{equation}
and $B<2\times 10^{12}\mbox{ G }(L/3\times10^{41}\mbox{ erg s$^{-1}$})^{1/2}$ are obtained from equations~(\ref{eq:magb}) and (\ref{eq:edot}) by noting that the bolometric and X-ray luminosities are seen to decrease at times $<60\mbox{ d}$ to a lowest value of $\sim10^{41}\mbox{ erg s$^{-1}$}$ \cite{Raf19}. 

The above discussion rules out the possibility of a millisecond magnetar powering AT2018cow.  If $B>10^{14}\mbox{ G}$, then $t_{\rm sd}<20\mbox{ d}$ and $-\dot{\nu}>1\times10^{-4}\mbox{ Hz s$^{-1}$}$.  The latter in particular implies the QPO frequency would have decreased by more than 100~Hz in a span of just 10 days. A young pulsar with a Crab-like magnetic field of a few times $10^{12}$ G is only possible in this scenario if an additional source of energy were present early on. A low-B value cannot explain the high and evolving luminosity early on in the outburst.

\subsection{Neutron star accretion}

Alternatively, a remnant magnetar may be accreting fall-back matter from its birthing supernova explosion, and the observed engine activity is either accretion or spin-down powered. The rate of mass fall-back can be written (e.g., \cite{Metzger18})
\begin{equation}\label{eq:eq7}
\dot{M}(t) = \frac{2}{3}\frac{M_{\rm acc}}{t_{\rm acc}}\frac{1}{(1 + t/t_{\rm acc})^{5/3}},
\end{equation}
where $M_{\rm acc}$ is the total quantity of returning mass and $t_{\rm acc}\sim t_{\rm ff}\sim(G\overline{\rho})^{-1/2}$ is the characteristic fall-back time which depends on the mean density $\overline{\rho}$ of the layer of progenitor star contributing to $M_{\rm acc}$. For an extended star with hydrogen-rich ejecta like a blue supergiant ($\bar{\rho} \sim 10^{-5}-10^{-3}$ g cm$^{-3}$), we have $t_{\rm acc}\sim10^{5}-10^{6}$~s, compatible with the timescale of engine activity for AT2018cow \cite{Raf19}. A neutron star of initial mass $M_{\rm ns}\approx1.4M_{\odot}$ can accrete at most $M_{\rm acc}\approx0.8M_{\odot}$ before collapsing into a black hole and thus indicates a maximum accretion rate in the range $\lesssim10^{-6}-10^{-5}M_{\odot}$~s$^{-1}$, i.e., 9-10 orders of magnitude above the Eddington accretion rate for a solar-mass compact object. By comparison, to explain the peak engine luminosity of AT2018cow ($\sim$10$^{45}$ erg s$^{-1}$ or 7 orders of magnitude above the Eddington luminosity) through accretion would require either a lower peak accretion rate $\dot{M}\sim10^{-8}M_{\odot}$~s$^{-1}$ or inefficient production of X-rays by the accretion flow. On timescales of several weeks, relevant to the epoch of the observed QPO, the accretion rate is $\sim$ 2 orders of magnitude lower than at peak, i.e. in the characteristic range
\begin{equation}
\dot{M}(t_{\rm QPO}) \sim 10^{-10} - 10^{-8} M_{\odot} {\rm s^{-1}} \label{eq:Mdotest}
\end{equation}

The Alfv\'en radius of the accretion flow, at which the ram pressure of the incoming flow is balance by magnetic forces, is given by (e.g., \cite{Metzger18})
\begin{equation}
R_{A} \approx 40 B_{12}^{4/7} \dot{M}_{-9}^{-2/7} M_{1.4}^{-1/7}\, {\rm km},
\end{equation}
where $\dot{M}_{\rm -9}$ = $\dot{M}$/(10$^{-9}$ M$_{\odot}$ s$^{-1}$) and now we have normalized the surface magnetic field B$_{12}$ = B/10$^{12}$ G to a lower value more akin to a radio pulsar than a magnetar. If the magnetic field is sufficiently large that $R_{\rm A}$ exceeds the neutron star radius $\approx$ 12 km, i.e. if
\begin{equation}
B_{12} \gtrsim 0.1\dot{M}_{-9}^{1/2} M_{1.4}^{1/4},
\end{equation}
then the accretion flow will be directed from the disk onto the magnetic axis of the neutron star, providing a possible mechanism to induce periodicity in the signal on the rotation period. 

In order for accretion to occur, a neutron star must rotate slow enough that it is not in an ejector phase \cite{Shvartsman71,Illarionov75,Lipunov92}, and this occurs approximately when the Alfv\'en radius $R_{\rm A}$ is smaller than the light cylinder radius $R_{\rm lc}\equiv c/2\pi\nu$.  This then implies
\begin{equation}
B_{12} < 20\,\dot{M}_{-9}^{1/2}M_{1.4}^{1/4}\left(\frac{\nu}{224\mbox{ Hz}}\right)^{-7/4}.
\end{equation}

Accretion also affects the spin-down rate of the neutron star, driving the spin-period to an equilibrium value given by \cite{Davidson73,Illarionov75,Piro11,Metzger18}$\footnote{Unlike in some past work, we assume that this equilibrium is determined by the balance of accretion spin-up and spin-down from the magnetized wind (the latter enhanced as a result of the larger magnetosphere opened by the fall-back \cite{parf16}).}$
\begin{equation}
P_{\rm eq} \approx 4\,B_{12}^{6/7} \dot{M}_{-9}^{-3/7} M_{1.4}^{-5/7}{\rm~ms} \label{eq:Peq}
\end{equation}
Explaining the observed QPO (P$_{\rm eq} \approx 1-2P_{\rm QPO} \approx 4.4-8.8$ ms) thus requires an accretion rate of
\begin{equation}
\dot{M}(t_{\rm QPO}) \approx (2-9)\times10^{-10} B_{12}^2 M_{1.4}^{-5/3} M_{\odot} {\rm s^{-1}},
\end{equation}
consistent with our estimate in eq.~\ref{eq:Mdotest}.  However the timescale to reach this equilibrium (Ref.~\cite{Metzger18}, their eq. 21;
see also \cite{Davidson73,Illarionov75,Lipunov92,Alpar01}),
\begin{equation}
\tau_{eq} \approx 400 {\rm d}~ B_{12}^{-8/7} \dot{M}_{-9}^{-3/7} M_{1.4}^{16/7} \underset{\dot{M} = \dot{M}_{\rm QPO}}{\approx}  (400-900)B_{12}^{-2}M_{1.4}^{16/7}\,{\rm d}
\end{equation}
can be long.  This indicates that $P \approx P_{\rm eq}$ will only be achieved on timescales of the observed QPO ($\tau_{eq} \ll 25 $d) if $B$ (or, equivalently, $\dot{M} \propto B^2$) is sufficiently high, $B_{12} \gtrsim 3$. In such a case that $\tau_{\rm eq} \gg t_{\rm QPO}$, then the spin period would reflect that of the neutron star at birth rather than $P_{eq}$, the equilibrium value achieved through accretion.

However, a direct connection between the above accretion scenario and
the observed behavior of AT2018cow is ruled out, for reasons that are
analogous to those which rule out an isolated magnetar.  In particular,
the time-dependence of luminosity and spin period predicted by a mass
accretion rate such as equation~(\ref{eq:eq7}) lead to constant
luminosity and spin period at times $t\ll t_{\rm acc}$ and evolving
luminosity and spin when $t\gg t_{\rm acc}$.  When applied to
AT2018cow, the observed declines of its X-ray and bolometric
luminosities \cite{Raf19} would imply a corresponding change in spin
period.  But such a change contradicts the constant frequency of the
QPO.

\section{Pulsation search assuming a binary}
To search for possible orbital periodicities, we used the open source PRESTO software\footnote{https://github.com/scottransom/presto} to perform an acceleration search over the frequency-frequency derivative plane. The acceleration search scheme assumes that the compact object’s acceleration is roughly constant throughout the observation of duration $ T \lesssim P_{\rm orb}/10$ \cite{ransom2002}. We ran the search assuming that any possible signal would drift across a maximum of 100 Fourier frequency bins (\emph{zmax} of 100). Informed by the centroid frequency and width of the QPO (assumed to be the fundamental frequency) reported above, we narrowed the search frequency window to be over 200-260 Hz\footnote{in PRESTO parlance, this is \emph{flo} of 200, and \emph{fhi} of 260}. The acceleration search yielded no candidates above 3 sigma (single trial probability). Subsequently, we accounted for the possibility of linearly changing accelerations, and carried out a jerk search \cite{andersen18}, with PRESTO\footnote{In PRESTO parlance, this is \emph{wmax} of 300}. This opens the way to detect other more exotic systems like very compact, relativistic binary systems, and allows searches over longer observations, unlike in acceleration searches. The jerk search would also allow us to recover any lost signal from residual Doppler smearing in the previous acceleration searches \cite{bagchi13}. In the end, the jerk search also yielded no candidates above 3 sigma (single trial probability).

As described in section \ref{supsec:ns}, the spin of the potential neutron star will evolve in a manner that depends strongly on the actual underlying physical scenario. Exploring all these models is beyond the scope of this work. But we provide the barycenter corrected events as supplementary files so that any reader with access to a computer cluster may use it. A sample Python code to load all the events is also provided (load\_events.py).


\section{A rough estimate of number of FBOT QPOs detectable with {\it NICER}}\label{supsec:rates}
The signal-to-noise of a QPO (n$_{\sigma}$, single trial) in a power spectrum depends on the mean source count rate (S), background rate (B), fractional rms of the QPO (rms), the total exposure time (T) and the width of the QPO (W). These quantities are related as follows \cite{vanderklis}:
\[
n_{\sigma} = \frac{1}{2}\frac{S^2}{S+B}\sqrt{\frac{T}{W}}
\]
Using a deep exposure of 250 ks, a conservative mean background rate of 0.5 cps, and an rms (30\%) and width ($<$16 Hz) similar to AT2018cow's 224 Hz QPO, we can use the above equation to estimate the minimum mean source count rate of a future FBOT to detect a QPO at 5$\sigma$ (single trial). This value is roughly 1.25 cps which is a factor of 5 lower than the peak 0.25-2.5 keV count rate. This translates to the source being roughly a factor of 2 farther in distance than AT2018cow, i.e., about 100 Mpcs. Using the volumetric rate of AT2018cow-like objects, i.e., $<$0.1\% of core-collapse supernovae = 0.001$\times$10$^{-4}$ events yr$^{-1}$ Mpc$^{-3}$ \cite{ccrate1, ccrate2, ztffbots} we estimate a rate of $\sim$0.3 events per year within 100 Mpcs. In other words, {\it NICER} could detect QPOs from AT2018cow-like if they are within 100 Mpc ($z\lesssim$0.025) which happen once every few years.  This estimate would improve slightly if {\it NICER} observations are strategically planned during the ISS orbit night when the overall background is lower. Other factors like large angular distance between the target and the sun, bright earth, etc, would also reduce the background but those would ultimately depend on the location of the target on the sky. 


\clearpage
\vfill\eject


\clearpage
\vfill\eject


\clearpage
\vfill\eject


\clearpage
\vfill\eject


\clearpage
\vfill\eject

\clearpage
\vfill\eject


\clearpage
\vfill\eject


\clearpage
\vfill\eject


\clearpage
\vfill\eject


\clearpage
\vfill\eject

\clearpage
\vfill\eject


\noindent\textbf{Movie S1:} The top panel of the movie shows the evolution of the average PDS and the gradual improvement in the QPO signal at 224 Hz with accumulated exposure. The lower panel shows the corresponding long-term light curve. The shaded red rectangle in the top panel shows the location of the 224 Hz QPO. This suggests that the QPO is long-lived and present in majority of the observations. This is available as {\it Movie\_S1.gif}. 
\pagebreak


\begin{table}
\normalsize
    \centering
    \caption{Upper limits on the fractional rms value of QPOs at various integer harmonics of 224 Hz (see Fig. \ref{fig:rmsupp}).\\ }
    \label{tab:rmstab}
    \begin{tabular}{|C{0.25\textwidth}|C{0.2\textwidth}|C{0.2\textwidth}|C{0.2\textwidth}|}
        \hline \T

        \textbf{Frequency range} & \textbf{rms upper limit (coherence=5)} & \textbf{rms upper limit (coherence=10)} & \textbf{rms upper limit (coherence=20)}  \\ \hline  \T
        $\frac{1}{2}\times$(224$\pm$16) Hz  & 30 & 25 & 23 \\ \T
        $\frac{3}{2}\times$(224$\pm$16) Hz  & 41 & 32 & 27 \\ \T
        2$\times$(224$\pm$16) Hz & 42 & 36 & 30  \\
         &&& \\ \hline
    \end{tabular}
\end{table}

\end{document}